\documentclass[aps,groupedaddress
,superscriptaddress
,amsfonts,amsmath,floatfix,eqsecnum,nofootinbib
]{revtex4}

\usepackage[usenames]{color}

\usepackage{epsfig}
\usepackage{graphicx}
\usepackage{mathptmx}      
\usepackage{latexsym}     
\usepackage{bm}
\usepackage{amstext}\usepackage{array}

\usepackage{bbm}
\usepackage{amsmath}
\usepackage{stmaryrd}

\def\tr{{\rm tr}}

\newcommand{\cA}{{\mathcal{A}}}
\newcommand{\cH}{{\mathcal{H}}}
\newcommand{\cM}{{\mathcal{M}}}

\newcommand{\SU}{{\mathrm{SU}}}
\newcommand{\U}{{\mathrm{U}}}
\newcommand{\Eq}[1]{Eq.~(\ref{eq:#1})}

\newcommand{\va}{{\bm{a}}}
\newcommand{\vb}{{\bm{b}}}
\newcommand{\vh}{{\bm{h}}}
\newcommand{\vk}{{\bm{k}}}
\newcommand{\vv}{{\bm{v}}}
\newcommand{\vx}{{\bm{x}}}

\newcommand{\vq}{{\bm{q}}}

\newcommand{\vA}{{\bm{A}}}
\newcommand{\vB}{{\bm{B}}}
\newcommand{\vC}{{\bm{C}}}
\newcommand{\vJ}{{\bm{J}}}
\newcommand{\vH}{{\bm{H}}}
\newcommand{\vL}{{\bm{L}}}
\newcommand{\vQ}{{\bm{Q}}}
\newcommand{\vS}{{\bm{S}}}
\newcommand{\vX}{{\bm{X}}}

\newcommand{\vbeta}{{\bm{\beta}}}
\newcommand{\vlambda}{{\bm{\lambda}}}
\newcommand{\vphi}{{\bm{\phi}}}
\newcommand{\vpsi}{{\bm{\psi}}}
\newcommand{\vnabla}{{\bm{\nabla}}}

\newcommand{\vsigma}{{\bm{\sigma}}}
\newcommand{\ignore}[1]{}

\newcommand{\R}{\mathbb{R}}
\newcommand{\C}{\mathbb{C}}

\renewcommand{\Re}{{\mathrm{Re}}\,}
\renewcommand{\Im}{{\mathrm{Im}}\,}
\newcommand{\su}{{\mathrm{su}}}

\newcommand{\esp}[1]{{\langle #1 \rangle}}

\newcommand{\mlpar}{(}\newcommand{\mrpar}{)}

\newcommand{\bph}[2]{ \mlpar #1 , #2 \mrpar }  
\newcommand{\bPh}[2]{ \left\mlpar #1 , #2 \right\mrpar }  
\newcommand{\pp}{\pc}

\newcommand{\ph}[2]{ \bph{ #1 }{ #2 } }
\newcommand{\Ph}[2]{ \bPh{ #1 }{ #2 } }
\newcommand{\pc}[2]{ \{ #1 , #2 \} }
\newcommand{\pq}[2]{ \llbracket #1 , #2 \rrbracket }
\newcommand{\php}[2]{ \ph{ #1 }{ #2 }^\prime }

\newcommand{\hA}{{\tilde{A}}}
\newcommand{\hcA}{{\tilde{\mathcal{A}}}}
\newcommand{\hQ}{{\tilde{Q}}}

\newcommand{\unit}{{\mathbbm{1}}}

\newcommand{\Esp}[1]{{\langle\!\!\langle #1 \rangle\!\!\rangle}}

\begin{document}

\title{Canonical bracket in quantum-classical hybrid systems}

\author{V. Gil}
\affiliation{Departamento de F{\'\i}sica At{\'o}mica, Molecular y Nuclear,
  Universidad de Granada, E-18071 Granada, Spain}
\author{L. L. Salcedo}
\affiliation{Departamento de F{\'\i}sica At{\'o}mica, Molecular y Nuclear,
  Universidad de Granada, E-18071 Granada, Spain}
\affiliation{Instituto Carlos I de F{\'\i}sica Te\'orica y Computacional,
  Universidad de Granada, E-18071 Granada, Spain}

\date{\today}

\begin{abstract}
We study compound systems with a classical sector and a quantum sector.  Among
other consistency conditions we require a canonical structure, that is, a Lie
bracket for the dynamical evolution of hybrid observables in the Heisenberg
picture, interpolating between the Poisson bracket and the commutator.  Weak
and strong postulates are proposed. We explicitly construct one such hybrid
bracket when the Hilbert space of the quantum sector is finite dimensional and
show that it is unique if the strong postulates are enforced. The adjoint
bracket for the Schrodinger picture version of the dynamics is also obtained.
Unfortunately, preservation of the positivity of the density matrix under the
evolution is not guaranteed. The case of a particle with classical position
and momentum and quantum spin-$\frac{1}{2}$ is discussed and the spin-orbit
dynamics is worked out.
\end{abstract}


\maketitle

\tableofcontents

\section{Introduction}
\label{sec:intro}

The purpose of this note is to consider hypothetical hybrid physical systems
involving a truly classical sector in interaction with a truly quantum sector,
and analyze the internal consistency of such constructions.  There are several
motivations in the literature to study those hybrid systems.

One motivation comes from the foundations and interpretation of quantum
mechanics. Classical and quantum sectors coexist in the Copenhagen
interpretation, where a classical measuring apparatus is postulated
\cite{Bohr:1958,Heisenberg:1958,dEspagnat:1976,Allahverdyan:2011cx}. A precise
formulation of this interpretation seems to require some consistent
description of quantum and classical sectors in mutual interaction.  Another
motivation, also based on the assumption of the existence of true classical
systems in nature, comes from Einstein gravity, as some authors have argued
that for all we know at present, gravity could be classical. For a discussion
see
\cite{Rosenfeld:1963,Kibble:1979jn,Isham:1980sb,Page:1981aj,Alvarez:1988tb,Boughn:2008jx,Carlip:2008zf,Derakhshani:2013qya}.

A totally different motivation applies for the more plausible point of view
that strictly speaking only quantum systems exist in nature and the classical
description is just a macroscopic approximation. Even so, there are many
situations where a fully quantum treatment is too difficult to work out and a
classical approximation is indicated for some subset of degrees of freedom. In
this regard a major impulse to the study of hybrid quantum-classical systems
comes from quantum chemistry, molecular physics or condensed matter physics,
typically using a quantum description for the fast degrees of freedom and a
classical description for the slow ones
\cite{Aqvist:1993,Monard:1996,Bornemann:1996,Prezhdo:1997,Kapral:1999,Csanyi:2004}. In
this point of view, the quantum-classical system would be intrinsically
approximated as a description of nature, yet it would be of interest to have
an internally consistent description of it. The situation would be analogous
to the problem of solving a set of ordinary differential equations using a
numerical method, thereby introducing some error; still one would like to have
some properties exactly preserved, as e.g., energy conservation or some exact
symmetries \cite{Magnus:1954zz,Blanes:2009}.

While classical mechanics is not a faithful description of nature, it is
nevertheless internally consistent. The case of mathematical consistency is
not quite so simple for hybrid quantum-classical systems, except in the
trivial case in which the two sectors are never coupled. Prescriptions are
needed to specify the hybrid dynamics, and many different approaches can be
found in the literature (see \cite{Barcelo:2012ja} for a classification of
them). Here we follow the path of previous works
\cite{Aleksandrov:1981,Boucher:1988ua,Anderson:1994si,Caro:1998us} which keep
a classical description for the classical sector and a quantum description for
the quantum sector, and emphasize the features that are common to quantum and
classical dynamics. Both in classical and quantum dynamics, the observables of
compound systems are sums of products of observables of each sector; we assume
the same in the hybrid case.  Another common feature is the canonical
structure. The main goal will be to construct a dynamical bracket for the
hybrid system, paralleling the Poisson bracket and the commutator, for the
evolution of the observables.  As it is well-known these are Lie brackets as
this property is needed to implement any canonical dynamics
\cite{Goldstein:2002}, so we seek the same property in the hybrid case.  Here
we study how far one could go in this direction.

We are guided by the following consideration. Let us consider two classical
sectors, each with its own phase space and with dynamics driven by the Poisson
bracket in each sector. In this case there is a standard way to combine the
two sectors into a new one: the phase space of the compound system is the
direct product of spaces (this is a fancy name for the set of linear
combinations of products of observables of the two sectors) and the dynamics
uses the Poisson bracket of the compound system. Instead of assuming this
dynamical bracket for the compound system from the outset, it can be asked
whether there are other possibilities. In \cite{Caro:1998us} it is shown that,
the (compound) Poisson bracket is the only solution if sufficiently
informative constraints are set. The analogous constraints also uniquely
select the commutator as the dynamical bracket of the compound system in the
case of two quantum sectors. (In passing, the direct product construction
guarantees that observables of two different quantum sectors commute.)  So for
hybrid quantum-classical systems we proceed in a completely similar manner,
that is, by taking the space of hybrid observables as the direct product space
and postulating conditions common to the classical-classical and the
quantum-quantum cases to be fulfilled by the hybrid dynamical bracket.

In Sec. \ref{sec:2} we give precise definitions of our set of hybrid
observables and specify how to obtain expectation values from them. Also we
lay down some general requirements to be satisfied by the hybrid dynamical
bracket in the Heisenberg picture. The basic requirement being, in addition to
a canonical structure, that each sector must evolve according to its own
dynamics when the coupling between them is switched off. Other conditions
cannot be maintained: Antisymmetry of the bracket already leads to an
incompatibility with the naive expectation of the bracket being a
derivation, so this latter requirement is dropped. However, the most
stringent condition is the requirement of the Jacobi identity, due to its lack
of linearity. For instance, considering two quantum sectors and taking their
classical limit, one recovers the classical dynamics which certainly fulfills
the Jacobi identity, but if the classical limit is taken in just one of the
sectors this results in a violation of the identity \cite{Caro:1998us}. 
(Classical-quantum limits have been studied in \cite{Oliynyk:2016jwt}.) As the
minimal postulates are too weak to fully solve for the set of brackets
satisfying them, we consider stronger (more restrictive) axioms, still
consistent with the correct behavior when the two sectors are decoupled. In
the same section we comment on alternative formulations found in the
literature which would lead more exotic hybrid scenarios.

In Sec. \ref{sec:3} we analyze the example of a particle with a classical
position and momentum and a quantum spin-$\frac{1}{2}$. We give a precise
formulation of such hybrid system showing that rotational invariance is
correctly realized on the system. We consider a concrete dynamics of the
spin-orbit type and solve explicitly the equations of motion in that case.

In Sec. \ref{sec:4} we explictly construct a hybrid dynamical bracket for the
particular case of quantum sectors with a {\em finite-dimensional} Hilbert
space. We show that this bracket fulfills the strong postulates laid down in
Sec. \ref{sec:2} and also that it is a Lie bracket, enjoying the Jacobi
identity property. This is very remarkable as it was proved in
\cite{Caro:1998us} that the strong postulates forbid any dynamical Lie bracket
when the quantum sector is of position-momentum type [i.e.,
  $\cH=L^2(\R^n)$]. To our knowledge, this is the first instance in the
literature of a dynamical hybrid bracket consistent with the Jacobi
identity. (We refer only to formulations where hybrid observables are linear
combinations of products of observables of each sector. Other formulations do
have Lie brackets for the hybrid dynamics, at the price of greatly enlarging
the set of admissible observables.)

Sec. \ref{sec:7} is rather technical and it is entirely devoted to prove that
the postulates yield precisely one hybrid bracket when the Hilbert space is
finite-dimensional. We constructively determine the most general solution to
the postulates and show that the solution is unique. The proof is similar to
that in \cite{Caro:1998us}, where a position-momentum quantum sector was
assumed and a no-go result was derived.

In Sec. \ref{sec:5} we discuss the Schrodinger picture for hybrid
systems. From the condition that the expectation values must coincide in
Heisenberg and Schrodinger pictures, we explicitly construct the adjoint
bracket for the dynamics in the Schrodinger picture. As it turns out such
bracket is not a Lie bracket nor a derivation.

In Sec. \ref{sec:6} we analyze whether the hybrid bracket previously obtained
complies with the essential requirement of maintaining the positivity of the
density matrix. We find that this is not guaranteed and show
counterexamples. This implies that the hybrid dynamics is not physically
meaningful, at least for generic hybrid systems. It is not excluded that
positivity could be preserved in certain particular cases. This negative
result could be a consequence of too strict postulates set on the hybrid
dynamics, or perhaps it could simply reflect that hybrid systems cannot be
formulated in a consistent manner as exact systems. In that case hybrid
dynamics would display intrinsic limitations, remaining as useful
approximations to fully quantum systems.

The conclusions are summarized in Sec. \ref{sec:8}.  In Appendix \ref{app:A}
we briefly consider the mathematical structure and possible
generalizations. In Appendix \ref{app:B} we provide the proof of a lemma
needing in Sec. \ref{sec:7}.

\section{The quantum-classical hybridization problem}
\label{sec:2}

\subsection{The space of observables}
\label{sec:2a}

Our starting point to formulate hybrid quantum-classical systems is the formal
similarity between classical and quantum mechanics within the canonical
formalism.

In the canonical formalism, a classical system is described by a phase space
with a set of coordinates $x^i,k_i$, $i=1,\ldots,n$ and the observables are
real functions $A(x,k)$ defined on the phase space. Let us denote $\cA_c$ the
set of classical observables. The dynamical evolution is given by\footnote{For
  simplicity we assume observables without explicit time dependence. Otherwise
  the equation would be $d A/dt = \pp{ A }{ H } + \partial A/\partial t$.}
\begin{equation}
\frac{d A}{dt} = \pp{ A }{ H } 
,
\label{eq:1.1}
\end{equation}
where $H(x,k)$ is the Hamiltonian and $\pp{~}{~}$ denotes the Poisson bracket
\begin{equation}
\pp{A}{B} = \sum_i \left(
\frac{\partial A}{\partial x^i}\frac{\partial B}{\partial k_i}
-
\frac{\partial A}{\partial k_i}\frac{\partial B}{\partial x^i}
\right)
.
\label{eq:2.2}
\end{equation}

Of course, the classical phase space needs not be of precisely this type, to
which we refer as {\em position-momentum} type in what follows. For instance,
one could work in the subset of observables depending only on the orbital
angular momentum $\vL = \vx \times \vk$, i.e., $A= f(\vL)$. These observables
close an algebra since the Poisson bracket can be worked out to give
\begin{equation}
\pp{A}{B} = \vL \cdot
\frac{\partial A}{\partial \vL} \times 
 \frac{\partial B}{\partial \vL}
,
\label{eq:2.3}
\end{equation}
and $\pp{A}{B}$ is again a function of $\vL$. Any set of structure constants
of a Lie algebra defines one such Poisson-Lie algebra where the observables
are functions of the generators regarded as classical variables.

We do not need to assume a classical sector of position-momentum type in our
discussion, for the classical sector it is only necessary that i) the
variables commute, the dynamical Poisson bracket be ii) a Lie bracket (i.e.,
linear, antisymmetric and fulfilling the Jacobi identity) and, iii) a
derivation (i.e., obeying the product law).\footnote{The word ``derivation''
  is used here in a technical sense. A linear application $D$ is a derivation
  with respect to a product $*$ if it fulfills the product law: $D(A*B) =
  D(A)*B+A*D(B)$. Throughout the paper ``derivation'' refers to the ordinary
  product, so for instance $D \equiv [A,~]$ (where $[~,~]$ is the commutator)
  is a derivation: $[A,BC]=[A,B]C+B[A,C]$. Exceptionally in Sec. \ref{sec:7}
  and Appendix \ref{app:A} ``derivation'' refers to a Lie product. For
  instance, the Jacobi identity establishes that $D \equiv (A,~)$ (being
  $(~,~)$ some Lie product) is a derivation with respect to the product
  $(~,~)$: $(A,(B,C))=((A,B),C)+(B,(A,C))$.}

In the quantum case, the states form a Hilbert space, and the observables
correspond to Hermitian operators. Let us denote $\cA_q$ the set of quantum
observables. In the Heisenberg picture they evolve as
\begin{equation}
\frac{d A}{dt} = \frac{1}{i\hbar}[A,H] 
=: \pq{A}{H}
,
\label{eq:1.4}
\end{equation}
where $[~,~]$ denotes the commutator. Like $\pp{~}{~}$ in $\cA_c$, the
operation $\pq{~}{~}$ is also a Lie bracket and a derivation in $\cA_q$. We
assume throughout that $\cA_q$ (the quantum observables) is the set of all
Hermitian operators of the quantum Hilbert space. In Appendix \ref{app:A} we
briefly consider relaxing this condition by demanding only that $\cA_q$ is a
Lie algebra, i.e., linear and closed under commutation, $\pq{\cA_q}{\cA_q}
\subseteq \cA_q$. (This is necessary if any observable can be part of the
Hamiltonian.)

For two classical particles with position and momentum $(\vx_1,\vk_1)$ and
$(\vx_2,\vk_2)$, the observables of the compound system are functions of
$(\vx_1,\vk_1,\vx_2,\vk_2)$. This is just the tensor product of the sets of
observables in both sectors, i.e., $\cA_c=\cA_c^{(1)}\otimes\cA_c^{(2)}$. A
similar tensor product occurs in the quantum case for the Hilbert space $\cH=
\cH^{(1)}\otimes\cH^{(2)}$ and hence for the set of observables $\cA_q =
\cA_q^{(1)}\otimes\cA_q^{(2)}$. For instance $\vS = \vS_1\otimes\unit_2 +
\unit_1 \otimes \vS_2$, for a two spin system. Quite naturally, we will assume
a similar construction for a hybrid quantum-classical system. That is, the set
of hybrid observables is identified with
\begin{equation}
\cA = \cA_c \otimes \cA_q
.
\label{eq:2.5}
\end{equation}
Alternative hybrid constructions in the literature not following this
structure are briefly discussed in Sec. \ref{sec:2c}.

According to this definition, the elements of $\cA$ are linear combinations of
products of c-number functions on the phase space with quantum operators.
Equivalently, a hybrid observable $A \in \cA$ will be any function defined on
the classical phase space taking values on the set of quantum operators
$\cA_q$.

Classical and quantum observables can be regarded as particular types of
hybrid observables, so $\cA_c$ and $\cA_q$ can be embedded within
$\cA$. $\cA_c$ can be identified with the hybrid observables of the type $C
\unit $ in $\cA$, where $\unit $ denotes the identity operator on $\cH$, and
$C$ is a c-number function of the classical variables. Likewise, $\cA_q$ can
be identified with the $Q \in \cA$ which are constant functions with respect
to the classical variables. Generic hybrid observables are linear combinations
of products $CQ$, with $C\in\cA_c$ and $Q\in\cA_q$. The $C$'s commute with
every observable, while $Q$'s commute with $C$'s but not with another
$Q^\prime$ in general. It would seem that the embeddings of $\cA_c$ and
$\cA_q$ in $\cA$ are asymmetric, as one requires being a multiple of $\unit$
and the other invokes the requirement of being a constant function. The
unifying criterion is that $\pq{A}{~}\equiv 0$ whenever $A\in\cA_c$ and
$\pc{A}{~}\equiv 0$ whenever $A\in\cA_q$. Constant multiples of $\unit$ are
simultaneously elements of $\cA_c$ and $\cA_q$.

The obvious way to assign expectation values to the hybrids observables is
\begin{equation}
\esp{A}_\rho =
\int d^nx \, d^nk \, \tr\big( \rho(x,k)A(x,k) \big)
,
\label{eq:2.6}
\end{equation}
where the integral is over the classical phase space and the trace is over the
quantum Hilbert space. The density matrix $\rho$ is a positive operator for
each $(x,k)$, with normalization $\esp{\unit }_\rho = 1$.  This form of
$\esp{A}_\rho$ interpolates between the purely classical and the purely
quantal cases.

By the same token the hybrid dynamics in $\cA$ requires a certain bracket
between observables, which will be denoted $\ph{~}{~}$, so that in the
Heisenberg picture
\begin{equation}
\frac{d A}{dt} = \ph{A}{H}
.
\label{eq:2.7a}
\end{equation}
The hybrid Schrodinger picture is discussed in Sec. \ref{sec:5}.

\subsection{Conditions on the hybrid bracket}
\label{sec:2b}

The hybrid dynamical bracket is subject to certain requirements. For instance,
if $C$ and $C^\prime$ denote classical observables and $Q$ and $Q^\prime$
denote quantum ones, one expects\footnote{As usual in the literature, when
  there is no possibility of confusion we omit the identity factor $\unit $
  that would appear, e.g., in $C\unit $.}
\begin{equation}
\ph{C}{C^\prime} = \pp{C}{C^\prime},
\quad
\ph{Q}{Q^\prime} = \pq{Q}{Q^\prime},
\quad
\ph{C}{Q} = 0,
\qquad
C,C^\prime \in \cA_c,
~
Q,Q^\prime \in \cA_q
.
\label{eq:1.8}
\end{equation}
This guarantees that when the two sectors are decoupled, that is, in the
particular case of a Hamiltonian of the type $H=C+Q$, these sectors behave
according to their usual dynamics. These conditions by no means fix the form
of $\ph{~}{~}$ since generic hybrid observables are linear combinations of
blocks $CQ$, with $C\in\cA_c$ and $Q\in\cA_q$, and a rule for
$\ph{CQ}{C^\prime Q^\prime}$ is needed.

Another obvious requirement is
\begin{equation}
\ph{A}{B}^\dagger = \ph{A}{B}
\qquad A,B\in\cA
,
\label{eq:1.8a}
\end{equation}
in order to ensure that Hermitian observables remain so under dynamical
evolution. Or more generally
\begin{equation}
\ph{A}{B}^\dagger = \ph{A^\dagger}{B^\dagger}
,
\end{equation}
if we include non necessarily Hermitian objects.\footnote{Genuine hybrid
  observables are Hermitian, however, for convenience we often extend the
  concept of observable to include complex functions and non Hermitian
  operators (more precisely, operators of the type $A+i B$ with $A$ and $B$
  Hermitian).}

Still another obvious requirement is that a constant observable, say $\unit$,
should not evolve. This would be automatically true {\em if $\ph{~}{~}$ were a
derivation} [see \Eq{2.18}]. In that case one would have
\begin{equation}
\ph{\unit}{H} = \ph{\unit^2}{H} = 2\ph{\unit}{H} = 0
\,.
\end{equation}
As we will show $\ph{~}{~}$ cannot be a derivation (at variance with 
$\pc{~}{~}$ and $\pq{~}{~}$), thus we postulate this property:
\begin{equation}
\ph{A}{\unit} = 0 \quad \forall A \in \cA
.
\label{eq:2.12}
\end{equation}

Of course, these considerations are not new. In \cite{Anderson:1994si} a
hybrid dynamical bracket was proposed, namely,
\begin{equation}
\bph{A}{B}_a \equiv \pq{A}{B} + \pp{A}{B}
.
\label{eq:1.9}
\end{equation}
Here $\pq{A}{B}$ is the same of \Eq{1.4} considering $A$ and $B$ as operators.
Also, $\pp{A}{B}$ is as defined exactly as in \Eq{2.2}, whether $A$ and $B$
are commuting or not. While the bracket in (\ref{eq:1.9}) is linear and obeys
the conditions (\ref{eq:1.8}), it is not antisymmetric. This latter condition
is needed to guarantee that, thanks to $\ph{H}{H}=0$, a time-independent
Hamiltonian is conservative also in the hybrid case. A more promising
prescription is \cite{Aleksandrov:1981,Boucher:1988ua}
\begin{equation}
\bph{A}{B}_s \equiv \pq{A}{B} +  \frac{1}{2}\big( \pp{A}{B} - \pp{B}{A} \big)
.
\label{eq:1.10}
\end{equation}
This is the hybrid bracket more frequently used in the
literature.\footnote{Actually, this bracket was introduced to describe the
  evolution of the density matrix, $d\rho/dt=\bph{H}{\rho}_s$, in the
  Schrodinger picture. The new insight of \cite{Anderson:1994si} was to
  introduce the hybrid bracket in the Heisenberg picture.} It is linear and
antisymmetric, however it fails to be a Lie bracket as the Jacobi identity is
not fulfilled. See \cite{Prezhdo:2006,Salcedo:2007zk,Prezhdo:2007} for other
similar brackets suffering the same problem.

Both the classical and the quantum brackets are Lie brackets, and this is also
an essential requirement to have a consistent canonical formalism in the
hybrid case.  The set of all invertible transformations of a system form a
group and the observables act as infinitesimal generators of it. The dynamical
evolution is generated by $H$ so that $\delta_H A = \ph{-Hdt}{A}$, but any
transformation can be chosen as the evolution. Thus other infinitesimal
transformations, such as symmetry groups, are generated through the same
bracket
\begin{equation}
\delta_G A = \ph{\delta G}{A}
, \qquad
\delta G= \sum_i \delta a^i G_i
,
\end{equation}
where the $a^i$ are the parameters of the group and the $G_i$ are the
generators. So for instance a relation
\begin{equation}
\ph{G}{H}=0
\end{equation}
is telling simultaneously that $\delta_HG=0$ and so $G$ is a constant of
motion, and that $\delta_G H=0$ and so $G$ generates a symmetry transformation
of the dynamics. Hence Noether theorem holds also in the hybrid case.  As
usual, that $A$ is a constant of motion means that $\esp{A}_\rho$ in any state
$\rho$ is conserved in time.

The Lie group structure of the transformations, and at the infinitesimal level
the Lie algebra structure, can only be represented by the action of
$\delta_G = \ph{\delta G}{~}$ as an operator in $\cA$ if $\ph{~}{~}$ is a Lie
bracket. Specifically, for two infinitesimal transformations generated by
$\delta_1G$ and $\delta_2G$, the commutator element\footnote{For two elements
  $g_1$, $g_2$ of a group, its commutator $g_{[1,2]}$ is defined as $g_1^{-1}
  g_2^{-1} g_1 g_2$.} is generated by $\delta_{[1,2]}G =
\ph{\delta_1G}{\delta_2G}$. The group structure demands $\delta_{[1,2]}
=[\delta_1,\delta_2]$ for any representation of the group
\cite{Schwinger:1970xc}, and so
\begin{equation}
\ph{
\ph{\delta_1G}{\delta_2G}
}{A}
=
\ph{\delta_1 G}{\ph{\delta_2 G}{A}}
-
\ph{\delta_2 G}{\ph{\delta_1 G}{A}}
.
\end{equation}
This is equivalent to the Jacobi identity, on account of the antisymmetry.

An immediate consequence of the Jacobi identity is that if the relation
$C=\ph{A}{B}$ holds at time $t=0$, $C(t)=\ph{A(t)}{B(t)}$ holds at any time
$t$, where $A(t)$ represents the Heisenberg picture evolution of $A$ and
similarly for $B$ and $C$.  More generally, $\delta_G = \ph{G}{~}$ is an
infinitesimal canonical transformation which preserves the canonical relations
between the canonical variables. Also, if $A$ and $B$ are constants of motion,
$\ph{A}{B}$ is again a constant of motion, thanks to the Jacobi identity.

Some times the generators of a group of transformations acting in a tensor
space are just the sum of generators in each factor space. For instance $\vJ =
\vL + \vS$ in $\cH= L^2(\R^3) \otimes \C^2$ for a quantum particle with spin
$1/2$. The orbital part generates the rotations in $L^2(\R^3)$ and
$\vS=\hbar\vsigma/2$ generates the rotations in $\C^2$. For $\cA = \cA_c
\otimes \cA_q$, one can consider generators of the type
\begin{equation}
G_i = C_i + Q_i,
\qquad
C_i\in \cA_c,
\quad
Q_i\in \cA_q
.
\label{eq:1.13}
\end{equation}
In this case, the conditions in \Eq{1.8} are already sufficient to guarantee
that $G_i$ reproduces the Lie algebra of the group,
\begin{equation}
\ph{G_i}{G_j} = c_{ij}{}^k G_k
,
\end{equation}
(where $c_{ij}{}^k$ are the structure constants) if $C_i$ and $Q_i$ fulfill
the analogous relations. Unfortunately, the tensor product (or rather tensor
sum) form in \Eq{1.13} is not sufficiently general. It is good enough for some
kinematical symmetry transformations, such as translations and rotations, or
some internal symmetries, but it already fails for the Poincare group, which
contains the boosts and the Hamiltonian within the generators. The Hamiltonian
takes the tensor sum form only in the rather trivial case of no interaction
between the classical and quantum sectors, and the same holds for the boost
generators.

Another question is whether the dynamical bracket should be a derivation, that
is, should fulfill Leibniz law,\footnote{Hence also $\ph{AB}{C} = A\ph{B}{C}
  + \ph{A}{C}B$ due to antisymmetry.}
\begin{equation}
\ph{A}{BC} = \ph{A}{B}C + B \ph{A}{C}
.
\label{eq:2.18}
\end{equation}
In fact the product law leads to a contradiction with linearity, antisymmetry
and $\ph{C}{Q}=0$~\cite{Salcedo:1995jr}. Consider for instance a hybrid system
with {\em classical} orbital angular momentum $\vL$ and {\em quantum} spin
$\vS$, then
\begin{equation}
0=\ph{\vL\cdot\vS}{\vL\cdot\vS}
=
\ph{L_iS_i}{L_j}S_j + L_j \ph{L_iS_i}{S_j}
=
\epsilon_{ijk}L_k S_i S_j + L_j L_i \epsilon_{ijk} S_k
= i\hbar\vL\cdot \vS
.
\end{equation} 
In the second equation we have used that $\vL$ and $\vS$ are angular momenta
[$\ph{L_i}{L_j}=\epsilon_{ijk}L_k$, and similarly $S_i$] and in the last
equality it has been used that $\vS\times\vS=i\hbar\vS$ but $\vL\times\vL=0$
classically, since the $L_i$ are commuting variables.

More generally, if there are two sectors such that $\ph{C}{Q}=0$ and $C$'s
commute with $Q$'s, and a bracket $\ph{~}{~}$ is a two-sided derivation, two
different expansions hold
\begin{equation}\begin{split}
\ph{CQ}{C^\prime Q^\prime} &= \ph{CQ}{C^\prime}  Q^\prime + C^\prime \ph{CQ}{
  Q^\prime}
= \ph{C}{C^\prime} QQ^\prime + C^\prime C \ph{Q}{ Q^\prime}
,
\\
\ph{CQ}{C^\prime Q^\prime} &=   \ph{C}{C^\prime Q^\prime} Q 
+ C \ph{Q}{C^\prime Q^\prime}
= \ph{C}{C^\prime} Q^\prime Q  + C  C^\prime \ph{Q}{ Q^\prime}
.
\label{eq:2.7}
\end{split}\end{equation}
We can see that there is a problem if the $C$'s are c-numbers and the $Q$'s
are not. More explicitly, subtracting the two expressions one finds
\begin{equation}
 \ph{C}{C^\prime} [Q, Q^\prime]  =  [C, C^\prime] \ph{Q}{ Q^\prime}
.
\end{equation}
This implies that such bracket is only consistent in two cases: i) all the
variables commute, i.e., both sectors are classical, or ii) $\ph{~}{~} =
[~,~]/(i\hbar)$ for some $\hbar$ {\em common to the two sectors}, i.e., both
sectors are quantal. The uniqueness of $\hbar$ has been noted in
\cite{Salcedo:1995jr,Sahoo:2004,Dass:2009qu}.

Therefore a hybrid quantum-classical bracket cannot be a derivation. The
product law would guarantee that, if $C=AB$ at $t=0$, $C(t)=A(t)B(t)$ at any
other time. The failure to comply with this implies that classical variables
can evolve into non commuting observables, in general.\footnote{In
  \cite{Salcedo:1995jr} it is argued that if observables of two sectors
  commute at $t=0$ but not at other times this implies an unacceptable
  privileged origin of time. In fact this is only so if the product is
  regarded as a privileged operation too: one can evolve the product at $t=0$
  into a new isomorphic $*$-product at time $t$ and observables would commute
  with respect to this new product. Going even further, one can change the
  representation of the observables so that in the new representation the
  $*$-product is just the ordinary product. This is the Schrodinger picture.}
This is the way quantum fluctuations propagate into the classical sector, the
so-called quantum back-reaction. In fact, as shown in \cite{Caro:1998us},
quantum fluctuations are not allowed for degrees of freedom which are purely
c-number variables.

It should be noted that the {\em product} of observables has a physical
meaning when the two observables commute with each other, namely, $AB$
represents the product of the numerical values of $A$ and $B$. This is the
case of two classical variables or of two observables belonging to two
different sectors. No such a physical meaning is attached to the product when
the observables do not commute. Actually, the product of two non-commuting
quantum Hermitian operators is not Hermitian, and so is not an
observable.\footnote{In some hybridization schemes the concept of product of
  two arbitrary hybrid observables is simply never introduced nor required.
  On the other hand a meaning can be given to the commutator $[A,B]$, since
  $\esp{-i[A(t_1),B(t_2)]}$ ~($t_1>t_2$) is the perturbation on $\esp{A(t_1)}$
  due to a small term $B$ added to the Hamiltonian at time $t_2$
  \cite{Negele:1988vy}.}

The axioms in Eqs. (\ref{eq:1.8}), (\ref{eq:1.8a}) and (\ref{eq:2.12}), plus
the Lie bracket condition are basic requirements on the hybrid dynamical
bracket. A further requirement is that positivity of a hybrid observable must
be preserved by the evolution. This is the case in quantum and classical
dynamics. We postpone the discussion of this requirement which is deferred to
Sec. \ref{sec:6}.

Regrettably the postulates in (\ref{eq:1.8}) are two general (too weak) to be
able to obtain the most general bracket consistent with them. So, to be able
to obtain definite answers, we will replace (\ref{eq:1.8}) with stronger
assumptions. Namely, following \cite{Caro:1998us} we will require
\begin{equation}
\ph{C}{A} = \pp{C}{A},
\quad
\ph{Q}{A} = \pq{Q}{A},
\qquad
\forall ~C\in\cA_c, ~Q\in\cA_q, ~A\in\cA
,
\label{eq:1.19}
\end{equation}
or equivalently
\begin{equation}
\ph{C}{C^\prime Q^\prime} = \pp{C}{C^\prime} \, Q^\prime,
\quad
\ph{Q}{C^\prime Q^\prime} = \pq{Q}{Q^\prime} \, C^\prime,
\qquad
C,C^\prime \in\cA_c, ~~Q,Q^\prime\in\cA_q
.
\end{equation}
Clearly these postulates imply those in \Eq{1.8}. They also guarantee
\Eq{2.12}, since $\unit$ is simultaneously a purely classical and a purely
quantal observable.

The justification of the stronger postulates is as follows: They are such that
the observable $C^\prime Q^\prime$ evolves into $C^\prime(t) Q^\prime$ if
$H=C$, and to $C^\prime Q^\prime(t)$ if $H=Q$. That is, if the two sectors are
dynamically uncoupled (i.e., $H=C+Q$) each factor in a product observable
$C^\prime Q^\prime$ evolves independently, according to its own dynamics. This
property certainly holds true for two classical sectors or for two quantum
sectors, so we postulate the same property in the hybrid case.

The axioms (\ref{eq:1.19}) immediately imply that the operators of the type
$\ph{C}{~}$ and $\ph{Q}{~}$, and hence also $\ph{C+Q}{~}$, act as derivations
in $\cA$. (This property will be used in Sec. \ref{sec:3}.)

On the other hand, $\ph{A}{~}$ is a derivation on $\cA_c$ or $\cA_q$ (but not
in $\cA$ in general), since
\begin{equation}
\ph{A}{CC^\prime} = \ph{A}{C}C^\prime + C\ph{A}{C^\prime},
\qquad
\ph{A}{QQ^\prime} = \ph{A}{Q}Q^\prime + Q\ph{A}{Q^\prime}
.
\end{equation}
An immediate consequence of this latter property is that when one or more
classical observables $C_i$ are constants of motion (i.e., $\ph{H}{C_i}=0$)
any function of them is also conserved. The same is true for functions of
quantum conserved observables. Nothing would be implied, in principle, for the
conservation of products of two arbitrary (hybrid) constants of motion.
Nevertheless, we show below that the product of a classical constant of motion
with a hybrid constant of motion is indeed conserved [point 7) at the end of
Sec. \ref{sec:4c}].

The axioms (\ref{eq:1.19}) do not directly stipulate any condition for
$\ph{CQ}{C^\prime Q^\prime}$, however, the Lie bracket requirement is rather
stringent. In fact, it is shown in \cite{Caro:1998us} that there is no Lie bracket
fulfilling (\ref{eq:1.19}) when both the classical and the quantum sectors are
of position-momentum type. The proof is made explicitly for the one
dimensional case but it holds for any number of dimensions. Also, the
precise nature of the classical sector is not relevant. This implies
that a fully consistent formulation of hybrid
quantum-classical systems with quantum sector $\cH=L^2(\R^n)$ does not exist,
at least if one insist on the stronger postulates (\ref{eq:1.19}).

What we show below is that if the Hilbert space of the quantum sector is
finite-dimensional, it is possible to construct a quantum-classical Lie
bracket obeying the axioms (\ref{eq:1.19}) plus (\ref{eq:1.8a}), and moreover,
such bracket is unique.\footnote{At least if one insist that $\cA_q$ is the
  set of all operators.} Unfortunately, we also show in Sec. \ref{sec:6} that
the hybrid bracket does not preserve positivity, and so it is not acceptable
as a consistent formulation.

\subsection{Alternative hybridization approaches}
\label{sec:2c}

Before leaving this section, we want to mention other approaches used in the
literature to define quantum-classical hybrid systems.

The traditional approach to mixing quantum and classical degrees of freedom is
through a mean-field or Ehrenfest dynamics. In such dynamics there is a
Hamiltonian $H(x,k)$ which is an Hermitian operator in the quantum Hilbert
space and a function of the classical variables, so it is a hybrid observable
of $\cA=\cA_c\otimes\cA_q$. The quantum sector evolves following the
Schrodinger equation, and there the classical degrees of freedom act as
time-dependent external parameters appearing in the quantum Hamiltonian. On
the other hand the classical degrees of freedom evolve classically using as
classical Hamiltonian the expectation value of the hybrid Hamiltonian in the
instantaneous quantum state. That is,
\begin{equation}
i\hbar\frac{d\psi}{dt} =  H(x,k)\psi
,\qquad
\frac{dx^i}{dt} = \frac{\partial\esp{H(x,k)}_\psi}{\partial k_i}
,\qquad
\frac{dk_i}{dt} = - \frac{\partial\esp{H(x,k)}_\psi}{\partial x^i}
.
\end{equation}
In this dynamics the two sectors affect each other but the classical variables
remain classical as they do not inherit any quantum fluctuations, hence the
so-called quantum back-reaction is missed \cite{Diosi:1999py,Craig:2005}.

Recently it has been emphasized that the mean-field dynamics can be expressed
in the form of a Lie bracket between hybrid observables
\cite{Zhang:2006,Alonso:2011,Elze:2011hi}. The observables are represented by
functions of $x^i$ and $k_i$ as well as $\psi_\alpha$ and
$i\hbar\psi^*_\alpha$, where $|\psi\rangle = \sum_\alpha \psi_\alpha
|\alpha\rangle$. For instance, any quantum operator $Q$ is represented by
$\esp{\psi|Q|\psi}$, and the Hamiltonian by $\esp{\psi|H(x,k)|\psi}$, while a
classical observable $C$ is represented by $C\esp{\psi|\psi}$. The
normalization $\esp{\psi|\psi}=1$ is preserved by the dynamics.  The dynamical
bracket is the Poisson bracket taking as $\psi_\alpha$ and
$i\hbar\psi_\alpha^*$ as canonical conjugated variables, in addition to $x^i$
and $k_i$. The dynamics is generated by the Poisson bracket between the
Hamiltonian and the observables, as in \Eq{1.1}.  From this point of view, the
canonical Ehrenfest dynamics has a proper Lie bracket and also preserves the
usual classical and quantum dynamics when the two sectors are not coupled in
the Hamiltonian.

There are several well-known problems with the mean field approach (see e.g.,
\cite{Boucher:1988ua,Diosi:1997mt,Craig:2005}). Here we highlight two of
them. 

The first problem is that initially one starts with observables of the type
$\esp{\psi|A|\psi}$ with $A \in \cA = \cA_c \otimes \cA_q$, but the Poisson
bracket [using $(x,k)$ and $(\psi,i\hbar\psi^*)$ as classical variables] does
not preserve this form (is not the expectation value of any operator).  So
$\ph{\cA}{\cA} \not\subseteq \cA$ and $\cA$ does not close a Lie algebra of
observables. The minimal algebra containing $\cA$ is very large, it probably
contains all kinds of functionals $F(x,k,\psi,i\hbar\psi^*)$ subjected only to
the condition of invariance under $(\psi,i\hbar\psi^*)\to
(\omega\psi,\omega^*i\hbar\psi^*)$ with $\omega \in \U(1)$, that is, arbitrary
functions of the blocks $\psi_\alpha^*\psi_\beta$. As argued above, in
classical or in quantum mechanics, when two different sectors are coupled, the
observables are linear combinations of products of the two sectors.  The
canonical Ehrenfest dynamics is qualitatively different in this regard, as it
prompts an explosion in the number of new hybrid observables. This is one
example of emergent phenomenon that would only arise when a classical and a
quantum sector are coupled.

The second problem is related to statistical consistency, as noted in
\cite{Salcedo:2012qx}: Let us assume that initially the two sectors are
decoupled and the hybrid state is a certain density matrix $\rho$ in the
quantum sector and some definite state in the classical sector (this is not
important), then let us switch on the coupling between both sectors. In order
to apply the mean-field dynamics one can proceed by decomposing $\rho$ as a
statistical mixture of (not necessarily orthogonal) pure states, that is,
\begin{equation}
\rho = \sum_n p_n |\varphi_n\rangle\langle \varphi_n|
,\quad \esp{\varphi_n|\varphi_n} = 1
,\qquad
p_n > 0, \quad \sum_n p_n = 1
.
\end{equation}
Then one can apply the mean-field evolution for each such pure state
$|\varphi_n\rangle$ coupled to the classical sector. By definition of
statistical mixture, the expectation values will be obtained by averaging over
the various histories (one for each pure state $|\varphi_n\rangle$) with
weight $p_n$.  However, the decomposition of a density matrix is not unique.
As it turns out \cite{Salcedo:2012qx} the averages of the hybrid observables
depend not only on $\rho$ but also on the concrete decomposition adopted. For
instance, a beam of unpolarized electrons (a quantum state) can be obtained by
an equal mixture of spin up and down electrons with respect to any given
quantization axis $\hat{n}$. According to the mean field dynamics, if the beam
passes through a chunk of {\em classical} material the evolution will depend
on the concrete choice of axis $\hat{n}$.  This would be another emergent
phenomenon: even if quantum mechanically two systems with the same density
matrix are completely indistinguishable, the coupling to the classical sector
would uncover hidden information in the mixture: the experiment would be able
to distinguish between different mixtures of pure states leading to the same
density matrix. Note that the presence of hidden information would be a
feature of ordinary quantum mechanics, even if it is only uncovered through
coupling to a classical sector. However, if all those new variables encoding
the hidden information in $\rho$ exists (in ordinary quantum mechanics) one
can ask whether they would not contribute to the entropy (since they encode
physical information) and get thermalized, as any other degree of freedom,
and reflect on the observed specific heat, for instance.

We find this very problematic. At variance with \cite{Barcelo:2012ja} we propose
here a {\em principle of non (strong) emergence} in quantum-classical hybrid
systems. If there were any truly classical systems in nature they would be the
measuring apparatuses of the Copenhagen interpretation. But it is precisely
through experiments that the standard quantum mechanics is established, thus
any quantum-classical hybrid approach implying extraordinary effects uncovered
when quantum systems are coupled to (hypothetical) classical ones should be
rejected, in our view.

Of course an easy way out would seem to express the mean field prescription
using a density matrix $\rho_q$ for the quantum sector:
\begin{equation}
i\hbar\frac{d\rho_q}{dt} =  [H(x,k),\rho_q]
,\qquad
\frac{dx^i}{dt} = \frac{\partial\esp{H(x,k)}_{\rho_q}}{\partial k_i}
,\qquad
\frac{dk_i}{dt} = - \frac{\partial\esp{H(x,k)}_{\rho_q}}{\partial x^i}
.
\label{eq:2.27a}
\end{equation}
However, even when a hybridization prescription is formulated using directly
$\rho_q$ (instead of pure states), statistical consistency is not
automatically guaranteed. By definition of mixture, if $\rho_q =
p_1\rho_1+p_2\rho_2$, we should demand for the expectation values that
$\esp{A}_{\rho_q} = p_1 \esp{A}_{\rho_1} + p_2 \esp{A}_{\rho_2}$.  Essentially
this requires linearity in $\rho_q$ in the evolution equations, and this
requirement is not fulfilled in \Eq{2.27a}.

Linearity holds true by construction in the hybrid Heisenberg picture
described above, \Eq{2.7a}. When the evolution of observables is transferred
to $\rho$, through \Eq{2.6}, one obtains in the Schrodinger picture
\begin{equation}
\frac{d \rho}{dt} = \php{H}{\rho}
,
\label{eq:2.7b}
\end{equation}
where $\php{~}{~}$ is the adjoint of $\ph{~}{~}$ (see Sec. \ref{sec:5}).  An
equation like (\ref{eq:2.7b}) cannot be written for the mean-field evolution.

It is easy to see that the two problems of explosion in the number of hybrid
observables and loss of statistical consistency are related and are intrinsic
to any mean-field--like dynamics. The idea behind the Ehrenfest formulation is
to describe the two sectors in the classical setting of the phase space $\cM$
with coordinates $\vX=(\vx,\vk,\vpsi,\vpsi^*)$, where $\vpsi$ is the vector
with components $\psi_\alpha$ (if desired one can use real vectors $\Re\vpsi$
and $\Im\vpsi$ as coordinates, this is unessential). In that dynamics each
point $\vX$ is a configuration of the hybrid system and it evolves under a
certain classical flow describing an orbit $\vX(t)$ on $\cM$. A mixed state
corresponds to a probability density $\sigma(\vX)$ on $\cM$.  If initially
$\sigma$ is Dirac delta at some point $\vX$, it will be a Dirac delta at
$\vX(t)$ at time $t$, so in particular the marginal density $\sigma(\vx,\vk)$
will also be a Dirac delta at all times: no fuzziness in the classical
variables is inherited by their coupling to the quantum sector, no quantum
back-reaction. The hybrid observables are functions $F(\vX)$ on $\cM$ which
evolve as $F(\vX(t))$.  The subset of observables in $\cA$ corresponds to
functions which are bilinear in $\vpsi$ and $\vpsi^*$, that is
$A_{\alpha\beta}(\vx,\vk) \psi_\alpha^* \psi_\beta$. If one goes outside this
set, new observables emerge and also statistical consistency is lost.  The
problem is that in all interesting cases $(\vx(t),\vk(t))$ will depend on the
initial $(\vpsi,\vpsi^*)$ and this additional dependence implies a breaking in
the bilinearity on $(\vpsi,\vpsi^*)$ in the observable. The problem would be
avoided if $(\vx(t),\vk(t))$ did not depend on the quantum sector but this
does not really describe two sectors in mutual interaction. The problem noted
is very general, even if the concrete flow of the mean-field dynamics were
replaced by something more sophisticated.

We could cast our own hybrid systems using the extended phase space $\cM$ and
observables as functions of the bilinear type $A_{\alpha\beta}(\vx,\vk)
\psi_\alpha^* \psi_\beta$ on it. We seek a suitable Lie bracket defined on
that set of functions but do not require from the outset that $A(X;t)=
A(X(t))$, which as we have just noted, would break the bilinearity condition.

There is a whole variety of hybridization prescriptions in the literature,
such as that based on the Hamilton-Jacobi approach
\cite{Hall:2005ax,Hall:2008}, the mapping of the classical sector into a
quantum one \cite{Sudarshan:1976bt,Chruscinski:2011}, or the Bohmian particle
\cite{Prezhdo:2001,Salcedo:2003mp,Prezhdo:2003}, and many other (see
e.g. \cite{Diosi:1999py,Radonjic:2012}), each with its own consistency
problems. Some of them are summarized in \cite{Barcelo:2012ja} and criticized
in \cite{Peres:2001,Caro:1998us,Salcedo:2012qx}. Here we stick to the
Heisenberg picture approach based on $\cA_c\otimes\cA_q$ with a canonical
formulation based on a dynamical Lie bracket, as it seems to be closest to a
standard purely quantum or purely classical description. This approach
bypasses the problem of explosion of hybrid observables, the emergence of
extraordinary effects uncovered by the hybrid coupling or problems with
statistical consistency. The one solution we find is flawed, however, by the
lack of positivity of the measure, a fatal problem already observed long ago
in \cite{Boucher:1988ua} for the bracket $\ph{~}{~}_s$.

\section{Particle with classical position-momentum and quantum spin-$\frac{1}{2}$}
\label{sec:3}

In this section we consider the hybrid system of a particle with classical
position and momentum and quantum spin-$\frac{1}{2}$.  Versions of this
problem have been considered before in the literature.  For instance,
Ref.~\cite{Schiller:1962} makes one such study in the context of a
Hamilton-Jacobi formulation, which is subject to the criticism expressed in
Sec. \ref{sec:2c}.

\subsection{Hybrid observables and bracket}

The observables of the classical sector are complex-valued functions
$f(\vx,\vk)$. This defines the set $\cA_c$. On the other hand, the quantum
Hilbert space is $\C^2$, so the quantum operators are $2 \times 2$ complex
matrices. This defines $\cA_q$. These can be expressed as $\sum_{\mu=0}^3
a_\mu \sigma_\mu$ where the $a_\mu$ are four complex numbers, $\sigma_i$,
$i=1,2,3$ are the Pauli matrices and $\sigma_0 = \unit $ is the $2 \times 2$
identity matrix.

The hybrid quantum-classical observables are generated from tensor product of
the quantum and classical sector, so they are linear combinations of products
of functions of $(\vx,\vk)$ with the $\sigma_\mu$. This defines
$\cA=\cA_c\otimes\cA_q$. Therefore, the hybrid observables are functions of
$\vx$ and $\vk$ that take values in the set of $2 \times 2$ matrices,
\begin{equation}
A\in\cA \quad
A(\vx,\vk)
= \begin{pmatrix} A_0+\frac{\hbar}{2}A_3 & \frac{\hbar}{2}A_1-\frac{i\hbar}{2}A_2 \\ 
 \frac{\hbar}{2}A_1+\frac{i\hbar}{2}A_2 & A_0-\frac{\hbar}{2}A_3 \end{pmatrix} 
= A_0(\vx,\vk) \unit  + \vA(\vx,\vk)\cdot\vS
,\qquad
\vS \equiv \frac{\hbar}{2}\vsigma
.
\end{equation}
For proper observables $A=A^\dagger$ and the functions $A_\mu(\vx,\vk)$ are
real.

A hybrid observable $A$ is purely classical when it acts as a multiple of the
identity on the quantum sector, i.e., when $A=A_0\unit $ and
$\vA\equiv 0$. Such observable commutes with any other observable, that is,
$\pq{A}{~}\equiv 0$.  On the other hand, $A$ is purely quantal when the four
functions $A_\mu(\vx,\vk)$ are actually independent of $(\vx,\vk)$ i.e., take
a constant value $a_\mu$ on the phase space. For such a purely quantal
observable $\pc{A}{~}\equiv 0$.

For a hybrid observable $A$, we will call $A_0$ the classical part of the
observable and $\vA\cdot\vS$ its quantal part.

To assign a canonical structure in the hybrid system, we need to define a Lie
bracket in the set of hybrid observables. A Lie bracket that fulfills the
axioms in \Eq{1.19}, as well as (\ref{eq:1.8a}), is the following one
\begin{equation}
\ph{ A_0 + \vA\cdot\vS
}{
B_0+\vB\cdot\vS
}
:=
\pp{ A_0 }{ B_0 }
+
\pp{ A_0 }{ \vB } \cdot \vS
+
\pp{ \vA }{ B_0 } \cdot \vS
+
\vA\times \vB \cdot \vS
.
\label{eq:3.3}
\end{equation}

It is easy to see that this bracket fulfills the axioms (\ref{eq:1.19}).
Indeed, when $B=C\unit $ is purely classical
\begin{equation}
\ph{A}{C} = 
\ph{
A_0 + \vA\cdot\vS
}{
C
}
=
\pc{ A_0 }{ C }
+
\pc{ \vA }{ C } \cdot \vS
=
\pc{A}{C}
,
\end{equation}
Likewise, when $Q=Q_0\unit  + \vQ\cdot\vS$ is purely quantal [the $Q_\mu$ are
constant with respect to $(x,k)$]
\begin{equation}
\ph{A}{Q} = 
\ph{A_0 + \vA\cdot\vS
}{
Q_0+\vQ\cdot\vS
}
=
\vA\times \vQ \cdot \vS
=
\pq{ A }{ Q }
.
\end{equation}

Obviously the bracket in \Eq{3.3} is also linear and antisymmetric.
Remarkably, it obeys the Jacobi identity too. This will follow as a particular
case in next section. In fact this is the unique Lie bracket consistent with
the postulates. This is shown in Sec. \ref{sec:7}.

The total angular momentum, with the usual form,
\begin{equation}
\vJ = \vL + \vS
,\qquad
\vL = \vx \times \vk
,
\end{equation}
fulfills the angular momentum algebra, $\ph{J_i}{J_j}=i\epsilon_{ijk}J_k$, and
is the generator of the rotations in $\cA$. As we noted in the previous
section, the postulates (\ref{eq:1.19}) guarantee that operators of the form
$\ph{C+Q}{~}$ are derivations. So $\ph{\vJ}{~}$ is a derivation and this is
useful to construct observables which are scalar under rotations, or more
generally, tensor with well defined transformations under rotations. For
instance, $\vk$ and $\vS$ are vectors and $\vk \cdot \vS$ is a scalar, i.e.,
\begin{equation}
\ph{J_i}{k_j}= \epsilon_{ijk} k_k
,\quad 
\ph{J_i}{S_j}= \epsilon_{ijk} S_k
,\quad 
\ph{\vJ}{\vk \cdot \vS}=0
.
\end{equation}
In general, a hybrid observable constructed with the basic vectors $\vx$,
$\vk$, and $\vS$ plus $\delta_{ij}$ and $\epsilon_{ijk}$, as usual, has the
standard tensorial behavior under rotations. This follows from the fact that
$\vL$ in $\vJ$ produces the usual rotation on the classical factors and $\vS$
on the quantum factors of the observable, while $\delta_{ij}$ and
$\epsilon_{ijk}$ are unchanged due to $\ph{\vJ}{\unit}\equiv 0$.

\subsection{Hybrid spin-orbit dynamics}

Let us consider a concrete dynamics for the particle with classical
position-momentum and quantum spin-$\frac{1}{2}$. For instance let
\begin{equation}
H = \frac{\vk^2}{2 M} +  g \vL\cdot\vS
.
\end{equation}
In this dynamics, the total angular momentum is conserved
\begin{equation}
\ph{\vJ}{H} = 0
,
\end{equation}
and in turn, $H$ is a scalar under rotations. Other constants of motion are
$\vk^2$, $\vL^2$, $\vL\cdot\vS$, and $\vS^2$  as is readily verified (in fact,
$\vS^2=\frac{3}{4}\hbar^2\unit$).

The classical orbital angular momentum and the quantum spin are not separately
conserved,
\begin{equation}
\ph{\vL}{H} = -\ph{\vS}{H} = -g \vL\times \vS
.
\end{equation}
This gives their variation to $O(t)$. The full evolution of $\vS$ is given by
\begin{equation}
\vS(t) = R^{-1}_t\vS,
\label{eq:3.10} 
\end{equation}
where $R_t$ is the rotation with parameters $\vphi = -gt\vL$.\footnote{For
  $\vphi=\phi\hat{n}$, with $\hat{n}^2=1$, $\hat{n}$ is the rotation axis and
  $\phi$ the rotation angle.} Correspondingly 
\begin{equation}
\vL(t) = \vJ - \vS(t) =
\vL - (R^{-1}_t-1)\vS,
\label{eq:3.10a} 
. 
\end{equation}
So $\vL(t)$ picks up a quantum component during its evolution. Note that
because the bracket is not a derivation $(AB)(t)$ needs not coincide with
$A(t)B(t)$. In particular $(\vL^2)(t)=\vL^2$ does not coincide with
$\vL(t)^2$. Nevertheless $\vL^2$ is a constant of motion and this is still a
meaningful non trivial statement, namely, the expectation value
$\esp{\vL^2}_{\rho(t)}$ remains time-independent in any state as $\rho(t)$
evolves in the Schrodinger picture.

Also, let us note that, even if $\dot\vL=\ph{\vL}{H}= -g \vL\times \vS$ (at
$t=0$), this does not imply that $\ddot\vL$ is given by $-g (\dot\vL \times
\vS + \vL \times \dot\vS)$, as the product law does not apply. Actually, that
expression is not even Hermitian. The correct result is instead
\begin{equation}
\ddot\vL = \ph{\ph{\vL}{H}}{H} = 
g^2(\vL^2 \, \vS - \vL\cdot\vS \, \vL)
.
\end{equation}

The full evolution of a generic observable can be worked out for
$H=g\vL\cdot\vS$, i.e., in the large $M$ limit. (Although in this case the
coupling $g$ can be eliminated by rescaling the time, we keep it for clarity.)
Let
\begin{equation}
A(\vx,\vk;t) = a(\vx,\vk;t) + \vb(\vx,\vk;t)\cdot \vS
.
\label{eq:3.13}
\end{equation}
The equations of motion (\ref{eq:2.7a}) can be written as
\begin{equation}
\frac{\partial a}{\partial t} = 0
,\qquad
\frac{\partial \vb}{\partial t} = g \pc{a}{\vL} + g \vb \times \vL
.
\end{equation}
Note that here $\vx$ and $\vk$, or $\vL\equiv \vx\times\vk$, are just
variables, they do not evolve. The evolution is on $a$ and $b$.

The first equation says that actually $a$ is time-independent. Taking a new
time derivative gives
\begin{equation}
\frac{\partial^2 \vb}{\partial t^2}  = g \frac{\partial \vb}{\partial t} 
\times \vL
,
\end{equation}
with solution
\begin{equation}
\frac{\partial \vb}{\partial t}  = R_t  \frac{\partial \vb }{\partial t}\Big|_{t=0}
,
\end{equation}
where $R_t $ the same as in \Eq{3.10} (i.e., a rotation of angle $-gtL$ over
the axis $\vL/L$). Thus the full solution is
\begin{equation}\begin{split}
a(t) &= a(0)
,\\
\vb(t) &= R_t \vb(0) + g t \frac{\vh\cdot\vL}{L^2} \vL
+\frac{1}{L^2} (R_t -1) (\vL\times\vh)
,
\label{eq:3.17}
\end{split}\end{equation}
where $L=|\vL|$ and
\begin{equation}\begin{split}
\vh \equiv  \pc{a}{\vL} = \vx\times\frac{\partial a}{\partial \vx}
+ \vk\times\frac{\partial a}{\partial \vk}
.
\end{split}\end{equation}

For $A(0)=\vS$ this formula reproduces \Eq{3.10}, and \Eq{3.10a} is reproduced
too. For $A(0)=\vx$ one obtains
\begin{equation}
\vx(t) = \vx
+ gt \vL \times \vx \frac{\vL\cdot\vS}{L^2}
+ \frac{\vL}{L^2} (R_t \vx-\vx) \cdot \vS
\,.
\end{equation}
The same expression can be written alternatively as
\begin{equation}
\vx(t) = \vx
+ gt \vS \times \vx
+ \frac{\vL}{L^2} (R_t \vx-\vx + gt \vL \times \vx) \cdot \vS
\,.
\end{equation}
In this form the connection with $\dot\vx(0) = \ph{\vx}{H} = g \vS \times \vx$
is more transparent since the last term is $O(t^2)$. Completely similar
expressions apply for $\vk(t)$.

Explicit solutions are also obtained for the special case in which $a$ and
$\vb$ are functions of $\vL$ only, as this property is preserved during
evolution with the spin-orbit Hamiltonian. This gives
\begin{equation}\begin{split}
a(t) &= a(0)
,\\
\vb(t) &= \vnabla a + R_t \left( \vb(0) - \vnabla a \right)
.
\label{eq:3.17b}
\end{split}\end{equation}
(The nablas refer a derivative with respect to $\vL$.) In this simpler
setting we have explicitly checked that the canonical bracket is preserved
during evolution, i.e., $\ph{A(t)}{B(t)} = \ph{A}{B}(t)$.

\section{Hybrid Lie bracket}
\label{sec:4}

In this section we show how hybrid brackets on $\cA$, fulfilling the
postulates of \Eq{1.19} and \Eq{1.8a}, can be constructed when the Hilbert
space of the quantum sector is finite dimensional,
\begin{equation}
\dim\cH = n < \infty
.
\end{equation}
In Appendix \ref{app:A} we comment on slightly more general constructions.

\subsection{Definition of the bracket}

To construct the bracket, let us define $\hcA_q\subseteq\cA_q$ as the set of
traceless operators,
\begin{equation}
\hcA_q = \{ \, \hQ \in \cA_q, ~ \tr(\hQ) = 0 \, \}
.
\end{equation}
Then, any $Q \in \cA_q$ can be uniquely decomposed as
\begin{equation}
Q  = Q_c + \hQ
\end{equation}
where $Q_c \propto \unit$ and $\tr(\hQ)=0$, with
\begin{equation}
Q_c \equiv \frac{1}{n}\tr(Q)\unit 
.
\end{equation}
In turn, the hybrid observables can be uniquely decomposed as
\begin{equation}
A  = C\unit + \hA
\end{equation}
where $C$ is classical and $\hA$ is traceless. So
\begin{equation}
\cA  = \cA_c \oplus \hcA,
\end{equation}
where $\hcA$ denotes the set of traceless hybrid
operators. This set is  $\hcA = \cA_c\otimes\hcA_q$, i.e.,
\begin{equation}
\hA =  \sum_i C_i \hQ_i
,
\qquad
C_i\in \cA_c
,\quad
\hQ_i \in \hcA_q
.
\end{equation}

The other property we need is that $\hcA_q$ is closed under commutation,
\begin{equation}
\pq{\hcA_q}{\hcA_q} \subseteq \hcA_q
.
\label{eq:4.3}
\end{equation}
This follows from $\tr([Q_1,Q_2]) = 0$ due to the cyclic property of the
trace.

In these definitions it is easy to recognize the pattern of the example of the
particle with classical position-momentum and quantum spin-$\frac{1}{2}$
considered in the previous section. There $\cA_c$ is $\{ A_0(\vx,\vk)\unit \}$
and $\hcA$ is $\{ \vA(\vx,\vk)\cdot \vS \}$.

Our statement is that a Lie bracket $\ph{~}{~}$ can be defined in $\cA$ as
follows:
\begin{equation}
\Ph{C + \hA }{ C^\prime + \hA^\prime } := 
\pc{C}{C^\prime}
+
\pc{\hA}{C^\prime}
+
\pc{C}{\hA^\prime}
+
\pq{\hA}{\hA^\prime}
.
\label{eq:4.6}
\end{equation}
Here $\pc{~}{~}$ and $\pq{~}{~}$ are the usual classical and quantum dynamical
brackets, and the $\unit$ operator is implicit.

More explicitly
\begin{equation}
\Ph{C \unit  + \sum_i C_i \hQ_i}{C^\prime \unit  + \sum_j C_j^\prime \hQ_j^\prime} = 
\pp{C}{C^\prime}\,\unit 
+
\sum_j \pp{C}{C_j^\prime} \, \hQ_j^\prime
+
\sum_i \pp{C_i}{C^\prime} \, \hQ_i
+
\sum_{i,j} C_i C_j^\prime \pq{\hQ_i}{\hQ_j^\prime}
.
\label{eq:4.8}
\end{equation}
Here, and also in the proof below, we use the symbols $\hQ$, $\hQ_i$, etc, to
denote operators in $\hcA_q$ (rather than to arbitrary elements of $\cA_q$).
The symbols $C$, $C_i$, etc, will denote elements of $\cA_c$, i.e., classical
observables, these are c-number functions on the phase space.

The definition in \Eq{3.3} for a particle with classical position-momentum and
quantum spin-$\frac{1}{2}$ is just an instance of the bracket in \Eq{4.8} for
the general case.

\subsection{Proof of the Lie bracket property}
\label{sec:4b}

Clearly, the construction in \Eq{4.6} is linear and antisymmetric and fulfills
the axioms \Eq{1.19}, as well as \Eq{1.8a}. We now show that it also fulfills
the Jacobi identity.  The Jacobi identity states that for any three hybrid
observables $A_1$, $A_2$ and $A_3$,
\begin{equation}
\ph{A_1}{\ph{A_2}{A_3}} + \ph{A_2}{\ph{A_3}{A_1}} 
+ \ph{A_3}{\ph{A_1}{A_2}} = 0
\label{eq:4.11}
\end{equation}
or equivalently (using antisymmetry)
\begin{equation}
\ph{A_1}{\ph{A_2}{A_3}} = \ph{\ph{A_1}{A_2}}{A_3} + \ph{A_2}{\ph{A_1}{A_3}} 
.
\end{equation}
For each observable it is sufficient to consider the two cases $A=C$ or
$A=C\hQ$.  The eight possibilities so generated can be classified by the number
$n_q$ of $\hQ$'s involved. It is only necessary to check the four cases
$n_q=0,1,2,3$.

Case $n_q=0$: For three classical observables $C_1$, $C_2$ and $C_3$
  (as usual we omit the factor $\bf{1}$)
\begin{equation}
\pp{C_1}{\pp{C_2}{C_3}} + \pp{C_2}{\pp{C_3}{C_1}} 
+ \pp{C_3}{\pp{C_1}{C_2}} = 0  
.
\end{equation}
This is trivially true being $\pp{~}{~}$ a Lie bracket.

Case $n_q=1$: For $A_1=C_1\hQ_1$ and $A_i=C_i$, $i=2,3$,
\begin{equation}\begin{split}
\ph{C_1 \hQ_1}{\ph{C_2}{C_3}} &= \pp{C_1}{\pp{C_2}{C_3} \,}  \,\hQ_1
\\
\ph{\ph{C_1\hQ_1}{C_2}}{C_3} + \ph{C_2}{\ph{C_1\hQ_1}{C_3}} &=
\ph{\pp{C_1}{C_2} \, \hQ_1}{C_3} + \ph{C_2}{\pp{C_1}{C_3} \, \hQ_1}
\\
&=\pp{\pp{C_1}{C_2}}{C_3} \, \hQ_1 + \pp{C_2}{\pp{C_1}{C_3}} \, \hQ_1
.
\end{split}\end{equation}
The two expressions coincide since $\pp{~}{~}$ is a Lie bracket.

Case $n_q=2$: For $A_1=C_1$ and $A_i=C_i\hQ_i$, $i=2,3$,
\begin{equation}
\ph{C_1 }{\ph{C_2 \hQ_2}{C_3 \hQ_3}} = \ph{C_1}{C_2 C_3 \pq{\hQ_2}{\hQ_3}}
= \pp{C_1}{C_2 C_3 }  \,\pq{\hQ_2}{\hQ_3}
,
\label{eq:4.13}
\end{equation}
using that $\pq{\hQ_2}{\hQ_3} \in \hcA_q$ ~[\Eq{4.3}] in the second equality.  On the
other hand
\begin{equation}\begin{split}
\ph{\ph{C_1}{C_2 \hQ_2}}{C_3 \hQ_3} + \ph{C_2 \hQ_2}{\ph{C_1}{C_3 \hQ_3}} &=
\ph{\pp{C_1}{C_2} \, \hQ_2}{C_3 \hQ_3} + \ph{C_2 \hQ_2}{\pp{C_1}{C_3} \, \hQ_3} 
\\
&= \pp{C_1}{C_2}  \, C_3 \pq{ \hQ_2}{ \hQ_3} + C_2 \pp{C_1}{C_3} \,\pq{ \hQ_2}{ \hQ_3} 
.
\label{eq:4.14}
\end{split}\end{equation}
The expressions in (\ref{eq:4.13}) and (\ref{eq:4.14}) coincide since
$\pp{~}{~}$ is a derivation and so $\pp{C_1}{C_2 C_3 } = 
\pp{C_1}{C_2} \, C_3 + C_2 \pp{C_1}{C_3} $.

Case $n_q=3$: For $A_i = C_i \hQ_i$, $i=1,2,3$,
\begin{equation}
\ph{C_i \hQ_i}{\ph{C_j \hQ_j}{C_k \hQ_k}} = \ph{C_i \hQ_i}{C_j C_k \pq{\hQ_j}{\hQ_k}}
=
C_i C_j C_k \pq{\hQ_i}{\pq{\hQ_j}{\hQ_k}}
.
\end{equation}
Clearly this expression vanishes upon adding cyclic permutations since this is
true for $\pq{~}{~}$ and the $C$'s commute.


This completes the proof.

\subsection{Observations}
\label{sec:4c}

The following observations are in order:

1) In the proof we have actually used only certain mathematical properties of
$\cA$ and $\hcA$. In Appendix \ref{app:A} we comment on more general
constructions relaxing the assumptions that $\cH$ must be finite dimensional,
$\cA_q$ should contain all operators and $\hcA_q$ should be the subset of
traceless operators in $\cA_q$.

2) We have noted that there is no hybrid dynamical Lie bracket fulfilling the
axioms in \Eq{1.19} when the classical and quantum sectors are of
position-momentum type \cite{Caro:1998us}.  The impediment comes from a
quantum sector with Hilbert space $\cH = L^2(\R^n)$. In this case the
condition $\pq{\hcA_q}{\hcA_q}\subseteq \hcA_q$ (used in the case $n_q=2$) is
violated since the commutator of two operators produces in general a component
along the $\unit$. An obvious example is that of the canonical commutation
relations $\pq{q^i}{p_j} = \delta^i_j \unit $.

3) The bracket in \Eq{4.6} is not a derivation. This can be seen by comparing
our prescription
\begin{equation}
\ph{C\hQ}{C^\prime \hQ^\prime} =  C C^\prime \pq{\hQ}{ \hQ^\prime}
.
\label{eq:4.16}
\end{equation}
with \Eq{2.7}, which has an extra term.

4) The bracket in \Eq{4.6} is somewhat similar to the ``standard'' proposal
$\bph{~}{~}_s$ in \Eq{1.10}, but bypasses the problem of the Jacobi identity by
treating separately the classical component ($C$ in $A=C+\hA$) and the quantum
component ($\hA$) of the observables. With the ``standard'' bracket one would
obtain the same result as with our Lie bracket for the cases covered by the
axioms, i.e., when one of the observables is purely classical or purely
quantal, while for the case of two properly hybrid observables
\begin{equation}
\bph{C\hQ}{C^\prime \hQ^\prime}_s =  C C^\prime \pq{\hQ}{ \hQ^\prime}
+ \frac{1}{2} (\hQ \hQ^\prime + \hQ^\prime \hQ) \pp{C}{C^\prime}
,
\end{equation}
which differs from our prescription in \Eq{4.16} by the last term.

5) It is easy to check that our bracket can also be expressed as
\begin{equation}
\ph{CQ}{C^\prime Q^\prime} 
=
C C^\prime \pq{Q}{Q^\prime} + \pp{C}{C^\prime}(Q Q^\prime - \hQ \hQ^\prime ) 
,
\end{equation}
where $Q = Q_c+\hQ$, ~ $Q^\prime = Q_c^\prime+\hQ^\prime$, with
$Q_c,Q_c^\prime \propto \unit$ and $\hQ, \hQ^\prime \in \hcA_q$, and
$C,C^\prime \in \cA_c$.  The factor $Q Q^\prime - \hQ \hQ^\prime$ is actually
symmetric with respect to $Q \leftrightarrow Q^\prime$.

6) Our bracket can be written in form similar to that in \Eq{3.3}, as follows.
Let $\{ q_i \}$, be a basis of $\hcA_q$ (similar to $S_i=\hbar\sigma_i/2$ for
spin-$\frac{1}{2}$) then, because $\hcA_q$ defines a Lie algebra with the
commutator,
\begin{equation}
\pq{q_i}{q_j} = c_{ij}{}^k q_k
,
\end{equation}
for some structure constants. The elements of $\hcA_q$ can be written as
$a = \va \cdot \vq = a^i q_i$ and so, using a standard notation,
\begin{equation}
\pq{a}{b} = \va \times \vb \cdot \vq
,\quad
(\va \times \vb)^k \equiv c_{ij}{}^k a^i b^j
.
\end{equation} 
With this notation our hybrid bracket takes the form
\begin{equation}
\ph{ A_0 + \vA\cdot\vq
}{
B_0+\vB\cdot\vq
}
=
\pp{ A_0 }{ B_0 }
+
\pp{ A_0 }{ \vB } \cdot \vq
+
\pp{ \vA }{ B_0 } \cdot \vq
+
\vA \times \vB \cdot \vq
.
\label{eq:3.3a}
\end{equation}

7) It is interesting to note that for this bracket the product of a {\em
  classical constant of motion} $C$ with a generic hybrid observable $A$,
gives
\begin{equation}
\ph{ C A }{ H } =  C \ph{ A }{ H } \quad \forall A\in\cA
.
\end{equation}
This follows from the fact that $\pc{~}{~}$ is a derivation and 
$\ph{C}{H_0} = \ph{C}{\vH}=0$, where $H=H_0+\vH\cdot\vq$.  An
immediate consequence is that if $A$ is also a constant of motion, the
product $CA$ is conserved too.  Of course, similar assertions hold when $C$ is
invariant under a transformation group.

\section{Uniqueness of the hybrid bracket}
\label{sec:7}

In this section we show that the hybrid bracket of \Eq{3.3a} is the only
solution to the double requirement of being a Lie bracket and fulfilling the
axioms in \Eq{1.19}.  This rather technical section is entirely devoted to
prove this.

We explicitly assume that the Hilbert space is finite-dimensional and $\cA_q$
is the set of all operators. The uniqueness does not automatically hold in the
generalizations discussed in Appendix \ref{app:A}.

Let $n=\dim\cH$, so $\cA_q$ is the set of $n \times n$ matrices. Any such
matrix can be written as
\begin{equation}
Q = a\unit + \vb \cdot \vlambda
,
\end{equation}
where the $\lambda_i$, $i=1,\ldots,n^2-1$ are the Gell-Mann matrices of
$\su(n)$ (a generalization of the Pauli matrices) \cite{Pascual:1984zb}. The
Gell-Mann matrices are $n^2-1$ linearly independent Hermitian and traceless
matrices which define a basis of $\hcA_q$:
\begin{equation}
  \lambda_i= \lambda_i^\dagger,\qquad
\tr(\lambda_i) = 0
.
\end{equation}
They fulfill the relations
\begin{equation}
\lambda_i \lambda_j  = \frac{2}{n}\delta_{ij} + d_{ijk}\lambda_k + i f_{ijk} \lambda_k
,
\label{eq:7.3}
\end{equation}
where the tensor $d_{ijk}$ is fully symmetric and $f_{ijk}$ is fully
antisymmetric. The $\lambda_i$ are normalized so that
$f_{ilm}f_{jlm}=n\delta_{ij}$.

The most general hybrid observable takes the form
\begin{equation}
A = C + \vC \cdot \vq
,
\end{equation}
where $C$ and $\vC$ are c-number functions on the phase space and we have
introduced the basis of $\hcA_q$
\begin{equation}
q_i \equiv \frac{\hbar}{2}\lambda_i
,
\end{equation}
which fulfills the commutation relations
\begin{equation}
\pq{q_i}{q_j} = f_{ijk} \, q_k
,
\end{equation}
as follows from \Eq{7.3}.

To fix the hybrid bracket we need to specify $\ph{C}{C^\prime}$,
$\ph{Cq_i}{C^\prime}$ and $\ph{Cq_i}{C^\prime q_j}$. (We already assume
antisymmetry of the bracket.) The first two are immediate from the
postulates,
\begin{equation}
\ph{C}{C^\prime} = \pc{C}{C^\prime}
,\qquad
\ph{C q_i}{C^\prime} = \pc{C}{C^\prime} \, q_i
,
\label{eq:7.7}
\end{equation}
and the postulates also require
\begin{equation}
\ph{C q_i}{q_j} = f_{ijk} C \, q_k
.
\label{eq:7.8}
\end{equation}
The most general form of $\ph{Cq_i}{C^\prime q_j}$ can be written as
\begin{equation}
\ph{C q_i}{C^\prime q_j} = \bph{C}{C^\prime}_{ij} \unit +
\bph{C}{C^\prime}_{ijk} \, q_k 
,
\end{equation}
where $\bph{C}{C^\prime}_{ij}$ and $\bph{C}{C^\prime}_{ijk}$ are two bilinear
operations to be specified. To do this let us enforce the Jacobi identity for
the three operators $C q_i$, $C^\prime q_j$ and $q_k$. On one hand
\begin{equation}
\ph{ \ph{C q_i}{C^\prime q_j} }{q_k} =
\bph{C}{C^\prime}_{ijl} f_{lkm} \, q_m
,
\end{equation}
while by Jacobi this equals
\begin{equation}
\ph{ \ph{C q_i}{q_k}}{C^\prime q_j}  + 
\ph{ C q_i}{\ph{C^\prime q_j}{q_k}} 
=
f_{ikl} \bph{C}{C^\prime}_{lj} + f_{jkl} \bph{C}{C^\prime}_{il}
+
f_{ikl} \bph{C}{C^\prime}_{ljm} \, q_m + f_{jkl} \bph{C}{C^\prime}_{ilm} \, q_m 
.
\end{equation}
This is equivalent to
\begin{equation}\begin{split}
0 &= f_{ikl} \bph{C}{C^\prime}_{lj} + f_{jkl} \bph{C}{C^\prime}_{il}
,
\\
0 &= f_{ikl} \bph{C}{C^\prime}_{ljm} + f_{jkl} \bph{C}{C^\prime}_{ilm} 
+ f_{mkl} \bph{C}{C^\prime}_{ijl}
.
\end{split}\end{equation}
The right-hand sides represent the infinitesimal $\SU(n)$ rotations of
$\bph{C}{C^\prime}_{ij}$ and $\bph{C}{C^\prime}_{ijm}$, and so the equations
imply that these are $\SU(n)$-invariant tensors. The only available invariant
tensors of rank $2$ and $3$ are those in \Eq{7.3}, thus this leads to
\begin{equation}
\ph{C q_i}{C^\prime q_j} = \bph{C}{C^\prime}_\delta \, \delta_{ij} \unit 
+ \bph{C}{C^\prime}_f f_{ijk} \, q_k 
+ \bph{C}{C^\prime}_d d_{ijk} \, q_k 
.
\label{eq:7.13}
\end{equation}
Here $\bph{~}{~}_\delta$, $\bph{~}{~}_f$ and $\bph{~}{~}_d$ are three auxiliary
brackets to be specified. The antisymmetry of $\ph{~}{~}$ requires
\begin{equation}\begin{split}
\bph{C}{C^\prime}_\delta &= - \bph{C^\prime}{C}_\delta
,\\
\bph{C}{C^\prime}_f &= + \bph{C^\prime}{C}_f
,\\
\bph{C}{C^\prime}_d &= - \bph{C^\prime}{C}_d
.
\end{split}
\label{eq:7.14}
\end{equation}
Further, \Eq{7.8} requires
\begin{equation}
\bph{C}{1}_\delta = \bph{C}{1}_d = 0 ,\qquad
\bph{C}{1}_f = C 
.
\label{eq:7.15}
\end{equation}

Of course, our hybrid bracket in \Eq{3.3a} fulfills all these
conditions. Specifically, for our hybrid bracket
\begin{equation}
\bph{C}{C^\prime}_\delta = \bph{C}{C^\prime}_d = 0 ,\qquad
\bph{C}{C^\prime}_f = C C^\prime 
.
\end{equation}
We want to show that this is actually the unique solution for a Lie bracket
fulfilling the constraints in Eqs. (\ref{eq:7.14}) and (\ref{eq:7.15}). 

To restrict the form of the auxiliary brackets $\bph{~}{~}_\delta$,
$\bph{~}{~}_f$, and $\bph{~}{~}_d$, we need to fully enforce the Jacobi identity
for three generic hybrid operators.  It is easy to verify that if there are
zero or one $q_i$ involved, Jacobi is fulfilled automatically. So the first
non trivial constraint comes from three operators with two $q_i$:
\begin{equation}
\ph{\ph{C q_i}{C^\prime q_j}}{C^{\prime\prime}}
=
\ph{\ph{C q_i}{C^{\prime\prime}}}{C^\prime q_j} +
\ph{C q_i}{\ph{C^\prime q_j}{C^{\prime\prime}}}
.
\end{equation}
Expanding this expression using Eqs. (\ref{eq:7.7}) and (\ref{eq:7.13}), and
noting that the three tensors $\delta_{ij}$, $f_{ijk}$ and $d_{ijk}$ are
linearly independent, leads to the relations
\begin{equation}\begin{split}
\pc{\bph{C}{C^\prime}_\delta}{C^{\prime\prime}}
&=
\bph{\pc{C}{C^{\prime\prime}}}{C^\prime}_\delta +
\bph{C}{\pc{C^\prime }{C^{\prime\prime}}}_\delta
,\\
\pc{\bph{C }{C^\prime }_f}{C^{\prime\prime}}
&=
\bph{\pc{C }{C^{\prime\prime}}}{C^\prime }_f +
\bph{C }{\pc{C^\prime }{C^{\prime\prime}}}_f
,\\
\pc{\bph{C }{C^\prime }_d}{C^{\prime\prime}}
&=
\bph{\pc{C }{C^{\prime\prime}}}{C^\prime }_d +
\bph{C }{\pc{C^\prime }{C^{\prime\prime}}}_d
.
\end{split}
\label{eq:7.18}
\end{equation}
In other words, the Poisson bracket acts as a derivation (in the sense of Lie
products \cite{Barut:1986dd}) with respect to the auxiliary brackets. In
Appendix \ref{app:B} we show that these conditions have as unique solutions
[using the symmetry conditions in \Eq{7.14}]\footnote{In the proof of Appendix
  \ref{app:B} we explicitly assume that the classical sector is of
  position-momentum type. It is not obvious whether the same proof covers more
  general cases.}
\begin{equation}
\bph{C}{C^\prime}_\delta = \alpha \pc{C}{C^\prime}
,\qquad
\bph{C}{C^\prime}_f = \beta C C^\prime
,\qquad
\bph{C}{C^\prime}_d = \gamma \pc{C}{C^\prime}
,
\label{eq:7.19}
\end{equation}
for some arbitrary constants $\alpha$, $\beta$ and $\gamma$. (Our bracket
corresponds to $\alpha=\gamma=0$ and $\beta=1$.)

The value $\beta=1$ follows from \Eq{7.15}. To fix $\alpha$ and $\gamma$ we
proceed to impose Jacobi with three $q_i$. Using previous results one obtains
(for convenience we retain an arbitrary $\beta$)
\begin{equation}\begin{split}
\ph{ C q_i }{ \ph{ C^\prime q_j }{ C^{\prime\prime} q_k} } 
&=
\alpha \beta \pc{ C }{ C^\prime C^{\prime\prime} } f_{ijk}
+
 \alpha \gamma \, \pc{ C }{ \pc{ C^\prime}{C^{\prime\prime} } }  \, d_{ijk}
+
\alpha \pc{C}{\pc{C^\prime}{C^{\prime\prime}}} \delta_{jk}\delta_{im}  \, q_m
\\
&~~~
+
 \beta^2 C C^\prime C^{\prime\prime} f_{jkl} f_{ilm}  \, q_m
+  \beta \gamma  \, \pc{ C }{ C^\prime C^{\prime\prime} } f_{jkl} d_{ilm}  \, q_m
+  \beta \gamma  \, C \pc{ C^\prime}{C^{\prime\prime} } d_{jkl} f_{ilm}  \, q_m
+
  \gamma^2 \pc{ C }{ C^\prime C^{\prime\prime} } d_{jkl} d_{ilm}  \, q_m
.
\end{split}\end{equation}
This expression must vanish upon summation on its three cyclic permutations.
Doing this, the terms with $\alpha \beta$, $\alpha \gamma$ and $\beta^2$
cancel identically. This leaves, for all $C$, $C^\prime$ and
$C^{\prime\prime}$, and all $i,j,k,m$,
\begin{equation}\begin{split}
0
&=
\alpha \pc{C}{\pc{C^\prime}{C^{\prime\prime}}} \delta_{jk}\delta_{im}  
+  \beta \gamma  \, \pc{ C }{ C^\prime C^{\prime\prime} } f_{jkl} d_{ilm} 
+  \beta \gamma  \, C \pc{ C^\prime}{C^{\prime\prime} } d_{jkl} f_{ilm}  
+
  \gamma^2 \pc{ C }{ C^\prime C^{\prime\prime} } d_{jkl} d_{ilm}  
 + \mathrm{c.p.}
,
\end{split}
\label{eq:7.21}
\end{equation}
where $\mathrm{c.p.}$ stands for cyclic permutations of
$(\{C,i\},\{C^\prime,j\},\{C^{\prime\prime},k\})$.

For $n \ge 4$ the rank-4 invariant tensors span a nine-dimensional
space,\footnote{This is just the number of times that the singlet
  representation appears in the direct product of four adjoint representations
  of $\SU(n)$.} and a basis is given by \cite{Pascual:1984zb}
\begin{equation}\begin{split}
&
\delta_{jk}\delta_{im}  ,\quad
\delta_{ki}\delta_{jm}  ,\quad
\delta_{ij}\delta_{km}  ,\quad
\\
&
d_{jkl} d_{ilm}  ,\quad
d_{kil} d_{jlm}  ,\quad
d_{ijl} d_{klm}  ,\quad
\\
&
d_{jkl} f_{ilm}  ,\quad
d_{kil} f_{jlm}  ,\quad
f_{ijl} d_{klm}  .
\end{split}\end{equation}
As a consequence, the vanishing of the components along
$\delta_{jk}\delta_{im}$ and $d_{jkl} d_{ilm}$ in \Eq{7.21} requires
\begin{equation}
0 = \alpha \pc{C}{\pc{C^\prime}{C^{\prime\prime}}} 
,\quad
0 = \gamma^2 \pc{ C }{ C^\prime
  C^{\prime\prime} }
.
\end{equation}
Since $\pc{C}{\pc{C^\prime}{C^{\prime\prime}}}$ or $\pc{ C }{ C^\prime
  C^{\prime\prime} }$ need not be zero, it follows that $\alpha=\gamma=0$,
which is our bracket (together with $\beta=1$).  For these values all terms in
\Eq{7.21} vanish as well, and this checks again that ours is a Lie bracket.

When $n=3$ the space of rank-4 invariant tensors span an eight-dimensional
space, due to the relation
\begin{equation}
\delta_{jk}\delta_{im} + \delta_{ki}\delta_{jm}  +
\delta_{ij}\delta_{km}  =
3(
d_{jkl} d_{ilm} +
d_{kil} d_{jlm}  +
d_{ijl} d_{klm} 
)
.
\end{equation}
Eliminating $\delta_{ij}\delta_{km}$, the vanishing of the components along
$\delta_{jk}\delta_{im}$ and $d_{jkl} d_{ilm}$  in \Eq{7.21} implies
\begin{equation}
0 = \alpha \left( \pc{C}{\pc{C^\prime}{C^{\prime\prime}}} 
-
\pc{C^{\prime\prime}}{\pc{C}{C^{\prime}}} 
\right)
,\quad
0 = \gamma^2 \, \pc{ C }{ C^\prime  C^{\prime\prime} }
- 3  \alpha \pc{C}{\pc{C^\prime}{C^{\prime\prime}}} 
.
\end{equation}
Once again the brackets need not vanish\footnote{For instance, for $C= x^2$,
  $C^\prime=k^2$ and $C^{\prime\prime}=xk$,~
  $\pc{C}{\pc{C^\prime}{C^{\prime\prime}}} -
  \pc{C^{\prime\prime}}{\pc{C}{C^{\prime}}} = -8xk$.  } so $\alpha=\gamma=0$.

For $n=2$, ~$d_{ijk}=0$ ($\bph{~}{~}_d$ does not appear) and the space has
dimension three with $\delta_{jk}\delta_{im}$, $\delta_{ki}\delta_{jm}$, and
$\delta_{ij}\delta_{km}$ as basis. \Eq{7.21} requires $\alpha=0$.

Throughout we have implicitly assumed $n>1$. When $\dim\cH = 1$ the $q_i$
do not exist. There is just one quantum state which is therefore completely
invariant under evolution or transformations. The dynamical bracket reduces to
the classical one.

This completes the proof of uniqueness.

It is interesting that a proof along the same lines can be provided for a
quantal-quantal bracket. This means the following: Let us have two quantum
systems provided with their usual dynamical bracket (the commutator), and let
us try to assign the most general dynamical bracket to the compound system
restricted by postulates similar to those imposed in the quantum-classical
hybrid case. Then one finds that the commutator is the unique solution as
a consistent dynamical bracket of the compound system. We have checked this when
one of the sectors is of position-momentum type and the other a
finite-dimensional Hilbert space (adapting the proof given in this
section). As shown in \cite{Caro:1998us} the same holds when the two quantum
sectors are of position-momentum type and also in a classical-classical
system, i.e., only the Poisson bracket is consistent as dynamical bracket of
the classical compound system.

\section{The Schrodinger picture}
\label{sec:5}

\subsection{Definition of the hybrid bracket in the Schrodinger picture}

Up to now we have been studying the evolution of hybrid systems within the
Heisenberg picture, in this section we obtain the dual description in the
Schrodinger picture, in which the density matrix $\rho$ carries the dynamical
evolution. To this end let us introduce the notation
\begin{equation}
\Esp{A}  
\equiv \int d^nx \, d^nk \, \tr \big( A(x,k) \big)
,
\label{eq:2.6a}
\end{equation}
that is, the symbol $\Esp{~}$ denotes integration over the classical phase
space plus trace over the quantum Hilbert space. In this way \Eq{2.6} can be
rewritten as
\begin{equation}
\esp{A}_\rho = \Esp{A \rho}
,
\label{eq:2.6b}
\end{equation}
where $\rho(x,k;t)$ is the density matrix of the hybrid system.

Throughout this section we assume that the Hilbert space is finite
dimensional, and as in the rest of the paper (except Appendix \ref{app:A}),
$\cA_q$ contains all operators and $\hA_q$ is the subset of traceless operators.
In this case
\begin{equation}
  \Esp{\hA} = 0 \quad \forall \hA \in \hcA
.
\end{equation}
Further properties are
\begin{equation}
\Esp{AB}=\Esp{BA}
,
\end{equation}
and also
\begin{equation}
\Esp{\pc{A}{B}} = \Esp{\pq{A}{B}} = 0
,
\label{eq:5.5}
\end{equation}
using integration by parts in the first case and the cyclic property of the
trace in the second case. More generally,
\begin{equation}
\Esp{\pc{A_1}{A_2}A_3} = \Esp{A_1\pc{A_2}{A_3}}
,\qquad
\Esp{\pq{A_1}{A_2}A_3} = \Esp{A_1\pq{A_2}{A_3}}
.
\label{eq:5.6a}
\end{equation}

In order to obtain the evolution of $\rho$, we postulate as usual that  the
expectation value must coincide in both pictures. Therefore
\begin{equation}
\frac{d}{dt}\esp{A}_\rho
= \Esp{ \frac{dA}{dt}\rho} = \Esp{A \frac{d\rho}{dt}}
,
\end{equation}
in the Heisenberg and Schrodinger pictures, respectively. As a consequence
\begin{equation}
\Esp{A \frac{d\rho}{dt}} 
= \Esp{\ph{A}{H} \rho}
.
\end{equation}
This holds for any $A$, and so $\rho$ obeys a certain linear equation which
can be written in the form
\begin{equation}
\frac{d \rho}{dt} = \php{H}{\rho}
.
\label{eq:5.9}
\end{equation}
Here $\php{A}{B}$ is an operation which is linear in $A$ and $B$.  The symbol
$\php{~}{~}$ is the dynamical bracket in the Schrodinger picture. From the
previous relations, it follows that this bracket satisfies the condition
\begin{equation}
\Esp{A\php{H}{\rho}} = \Esp{ \ph{A}{H}\rho}
.
\label{eq:5.6}
\end{equation}
Since $A$ is completely arbitrary, this relation fully fixes $\php{H}{\rho}$:
indeed, $\Esp{AB}=0$ for all $A$ can only hold if $B=0$, and so
$\Esp{AB_1}=\Esp{AB_2}$ for all $A$ implies $B_1=B_2$. Alternatively,
\begin{equation}
\frac{\delta \Esp{AB}}{\delta A} = B
,
\end{equation}
implies that $B$ is fixed from the knowledge of $\Esp{AB}$ for all $A$.

Let us note that when all observables are classical $\Esp{ \pc{A}{H}\rho} =
\Esp{A \pc{H}{\rho}}$ [using \Eq{5.6a}]. Likewise in the quantum case $\Esp{
  \pq{A}{H}\rho} = \Esp{A \pq{H}{\rho}}$. So in these two cases the bracket
takes the same form in the Heisenberg and Schrodinger pictures:
\begin{equation}
\pc{~}{~}^\prime=\pc{~}{~}
,\qquad
\pq{~}{~}^\prime=\pq{~}{~}
.
\end{equation}
However, in the hybrid case $\php{~}{~}$ needs not coincide with $\ph{~}{~}$,
and in fact it does not.

\subsection{Explicit construction of the hybrid bracket in the Schrodinger picture}

In order to fully determine the bracket from \Eq{5.6}, we will consider the
most general hybrid operator $A$, which is of the type $C + C^\prime \hQ$,
with $C,C^\prime\in\cA_c$ and $\hQ\in \hcA_q$ (that is, $\tr(\hQ)=0$). We follow
a procedure similar to that in the proof of the Jacobi identity, namely, we
workout the various cases classified by the number $n_q$ of $\hQ$'s present in
the operators $\rho$ and $H$ in \Eq{5.6}.

Case $n_q=0$: For arbitrary $C_i$ and $\hQ_i$
\begin{equation}\begin{split}
\Esp{C_1\php{C_2}{C_3}\,} &= \Esp{\ph{C_1}{C_2}C_3} = \Esp{C_1\pc{C_2}{C_3}},
\\
\Esp{C_1\hQ_1\php{C_2}{C_3}\,} &= \Esp{\ph{C_1\hQ_1}{C_2}C_3} 
= \Esp{C_1\hQ_1\pc{C_2}{C_3}},
,
\end{split}\end{equation}
where we have used \Eq{5.6a}.  Hence, for all $A$ ~
$\Esp{A\php{C_2}{C_3}\,}  = \Esp{A\pc{C_2}{C_3}}$ and so
\begin{equation}
\php{C_2}{C_3} =\pc{C_2}{C_3}
.
\end{equation}

Case $n_q=1$. Since we do not assume that $\php{~}{~}$ is antisymmetric, we
have to distinguish two cases, depending on whether the $\hQ$ is in $H$ or in
$\rho$. In the first instance
\begin{equation}\begin{split}
\Esp{C_1\php{C_2\hQ_2}{C_3}\,} &= 
\Esp{\ph{C_1}{C_2\hQ_2}C_3} = \Esp{C_1\pc{C_2}{C_3}\hQ_2} = 0,
\\
\Esp{C_1\hQ_1\php{C_2\hQ_2}{C_3}\,} &= 
\Esp{\ph{C_1\hQ_1}{C_2\hQ_2}C_3} = \Esp{C_1 C_2 C_3 \pq{\hQ_1}{\hQ_2}} = 0
.
\end{split}\end{equation}
Where it has been used that $\tr(\hQ) = \tr([A,B])=0$. Therefore
\begin{equation}
\php{C_2\hQ_2}{C_3} = 0
.
\end{equation}
On the other hand, when the factor $\hQ$ is in $\rho$
\begin{equation}\begin{split}
\Esp{C_1\php{C_2}{C_3\hQ_3}\,} &= 
\Esp{\ph{C_1}{C_2}C_3\hQ_3} = \Esp{C_1\pc{C_2}{C_3}\hQ_3} ,
\\
\Esp{C_1\hQ_1\php{C_2}{C_3\hQ_3}\,} &= 
\Esp{\ph{C_1\hQ_1}{C_2}C_3\hQ_3} = \Esp{C_1 \hQ_1 \pc{C_2}{C_3} \hQ_3} 
.
\end{split}\end{equation}
Hence,
\begin{equation}
\php{C_2}{C_3 \hQ_3} = \pc{C_2}{C_3}\hQ_3
.
\end{equation}

Case $n_q=2$. In this case we derive the following relations
\begin{equation}\begin{split}
\Esp{C_1\php{C_2\hQ_2}{C_3\hQ_3}\,} &= 
\Esp{\ph{C_1}{C_2\hQ_2}C_3\hQ_3} = \Esp{C_1\pc{C_2}{C_3}\hQ_2\hQ_3},
\\
\Esp{C_1\hQ_1\php{C_2\hQ_2}{C_3\hQ_3}\,} &= 
\Esp{\ph{C_1\hQ_1}{C_2\hQ_2}C_3\hQ_3} = \Esp{C_1 \hQ_1 C_2 C_3\pq{\hQ_2}{\hQ_3}}
.
\end{split}\end{equation}
The second relation implies
\begin{equation}
\php{C_2\hQ_2}{C_3\hQ_3}
= C_2 C_3\pq{\hQ_2}{\hQ_3}
+ C
\label{eq:5.21}
\end{equation}
where $C$ is some element of $\cA_c$. This ambiguity follows from
$\Esp{C_1\hQ_1 C} = 0$. On the other hand, from the first relation it
follows that
\begin{equation}
\php{C_2\hQ_2}{C_3\hQ_3}
= \pc{C_2}{C_3}\hQ_2\hQ_3 + \hA
\end{equation}
where $\hA$ is some undetermined element of $\hcA$, since $\Esp{C_1\hA}=0$.
To fulfill both conditions we rewrite this relation in the form
\begin{equation}
\php{C_2\hQ_2}{C_3\hQ_3}
= \pc{C_2}{C_3}\frac{1}{n}\tr(\hQ_2\hQ_3) + \hA^\prime
,
\label{eq:5.23}
\end{equation}
using that $\hQ_2\hQ_3 = \frac{1}{n}\tr(\hQ_2\hQ_3) + \hQ^\prime$, with
$\hQ^\prime$ traceless. Here $n= \dim\cH = \tr(\unit)$. The conditions in
Eqs. (\ref{eq:5.21}) and (\ref{eq:5.23}) are consistent identifying
$C$ with $\pc{C_2}{C_3}\frac{1}{n}\tr(\hQ_2\hQ_3)$ and 
$\hA^\prime$ with $C_2 C_3\pq{\hQ_2}{\hQ_3}$, so finally
\begin{equation}
\php{C_2\hQ_2}{C_3\hQ_3}
=  C_2 C_3\pq{\hQ_2}{\hQ_3} + \pc{C_2}{C_3}\frac{1}{n}\tr(\hQ_2\hQ_3)
.
\end{equation}

In summary, the dynamical bracket in the Schrodinger picture takes the form
\begin{equation}
\php{C_1 + C_1^\prime\hQ_1}{C_2 + C_2^\prime\hQ_2}
=
\pc{C_1}{C_2} + \pc{C_1}{C_2^\prime} \hQ_2 
+
  C_1^\prime C_2^\prime \pq{\hQ_1}{\hQ_2} 
+ \pc{C_1^\prime}{C_2^\prime}\frac{1}{n}\tr(\hQ_1\hQ_2)
.
\label{eq:5.24}
\end{equation}
This reduces to the classical or quantum brackets when the observables are
purely classical or purely quantal, respectively, as it is easily verified.

The density matrix must fulfill the conditions
\begin{equation}
\Esp{\rho} = 1,
\qquad
\rho^\dagger = \rho
.
\end{equation}
These relations are preserved by the evolution induced by the bracket since
\begin{equation}
\Esp{\php{H}{\rho}} \equiv 0
\end{equation}
using \Eq{5.5}, and also
\begin{equation}
\php{A^\dagger}{B^\dagger} = \php{A}{B}{}^\dagger
.
\end{equation}

The dynamical bracket in the Schrodinger picture is not a Lie bracket (is
not even antisymmetric). Also $\php{H}{~}$ is not a derivation. For instance,
for a derivation one would have (with $H= C_1^\prime\hQ_1$)
\begin{equation}
\php{C_1^\prime\hQ_1}{C_2^\prime \hQ_2}
=
\php{C_1^\prime\hQ_1}{C_2^\prime}  \hQ_2 + C_2^\prime \php{C_1^\prime\hQ_1}{\hQ_2}
=
0 +  C_2^\prime  C_1^\prime  \pq{\hQ_1}{\hQ_2} 
\end{equation}
while actually
\begin{equation}
\php{C_1^\prime\hQ_1}{C_2^\prime \hQ_2}
=
\pc{C_1^\prime}{C_2^\prime}\frac{1}{n}\tr(\hQ_1\hQ_2)
 +  C_2^\prime  C_1^\prime  \pq{\hQ_1}{\hQ_2} 
.
\end{equation}

The Schrodinger picture hybrid bracket can be written in a form similar to
\Eq{3.3a}, as 
\begin{equation}
\php{ A_0 + \vA\cdot\vq
}{
B_0+\vB\cdot\vq
}
=
\pp{ A_0 }{ B_0 }
+
\pp{ A_0 }{ \vB } \cdot \vq
+
\vA \times \vB \cdot \vq
+
\frac{\hbar^2}{2n}\pc{\vA\cdot}{\vB}
,
\label{eq:5.28}
\end{equation}
where, for $a=a^i q_i$ and $b= b^i q_i$, 
\begin{equation}
\va \cdot \vb \equiv g_{ij} a^i b^j
,\qquad
g_{ij} = \frac{2}{\hbar^2}\tr(q_iq_j)
.
\end{equation} 

\subsection{Classical particle with classical position-momentum and quantum spin-$\frac{1}{2}$}

For illustration purposes, let us workout the evolution of $\rho(t)$ for the
Hamiltonian $H=g\vL\cdot\vS$, discussed in Sec. \ref{sec:3}. The general form
of the density matrix is
\begin{equation}
\rho(\vx,\vk;t) =  \alpha(\vx,\vk;t)  + \vbeta(\vx,\vk;t) \cdot \vS
.
\end{equation}
The equation of motion (\ref{eq:5.9}) leads to
\begin{equation}
\frac{\partial \alpha}{\partial t} = \frac{\hbar^2}{4}g\pc{\vL\cdot}{\vbeta}
,\qquad
\frac{\partial \vbeta}{\partial t} = g \vL \times \vbeta
.
\end{equation}
These equations can be solved to give
\begin{equation}\begin{split}
\alpha(t) &= 
\alpha(0) -\frac{\hbar^2}{4L^2}\tr\left(R^{-1}_t - 1 \right)\vL\cdot\vbeta(0)
+
\frac{\hbar^2}{4L^2} \left(
gt L_i L_j -(R^{-1}_t - 1)_{il} \epsilon_{lmj}L_m
\right) h_{ij}
,
\\
\vbeta(t) &= R^{-1}_t \vbeta(0)
,
\label{eq:5.30}
\end{split}
\end{equation}
where $R_t $ is the same rotation as in Sec. \ref{sec:3}, and
\begin{equation}
h_{ij} \equiv \pc{L_i}{\beta_j(0)}
.
\end{equation}

The expectation value with an observable $A=a+\vb\cdot\vS$ takes the form
\begin{equation}
\Esp{A\rho} = \Esp{ a\alpha + \frac{\hbar^2}{4}\vb\cdot\vbeta}
.
\label{eq:6.36}
\end{equation}
The equivalence of the Heisenberg and Schrodinger pictures requires
\begin{equation}
\Esp{ a(t)\alpha(0) + \frac{\hbar^2}{4}\vb(t)\cdot\vbeta(0)}
=
\Esp{ a(0)\alpha(t) + \frac{\hbar^2}{4}\vb(0)\cdot\vbeta(t)}
.
\end{equation}
Using integration by parts [\Eq{5.6a}] one can verify, after some algebra,
that Eqs. (\ref{eq:3.17}) and (\ref{eq:5.30}) fulfill these relations. 

In the simpler setting of $\rho$ being a function of $\vL$ only (a property
preserved by the evolution with the spin-orbit dynamics) one obtains
\begin{equation}\begin{split}
\alpha(t) &= 
\alpha(0) + \frac{\hbar^2}{4} \vnabla \cdot (R_t^{-1} - 1 ) \vbeta(0)
,
\\
\vbeta(t) &= R^{-1}_t \vbeta(0)
.
\label{eq:5.30b}
\end{split}
\end{equation}
From \Eq{6.36} and the formula of $\alpha(t)$ it is readily verified that $
\esp{\vL^2}_{\rho(t)} = \Esp{ \vL^2\alpha(t)}$ is independent of $t$ for any
initial $\rho(0)$.

\section{The positivity fiasco}
\label{sec:6}

As we have just shown, rather general requirements seem to completely fix in a
unique way the form of the hybrid dynamics. In this section we show that
unfortunately the requirement of positivity is not fulfilled by such dynamics.
The same problem arises in the standard bracket of \Eq{1.10}
\cite{Boucher:1988ua}. Similar problems with positivity have been discussed in
\cite{Ahmadzadega:2016}.

\subsection{Heisenberg picture}

In the Heisenberg picture, the requirement of positivity is a particular case
of the more general requirement of the {\em preservation of the range} of an
observable.

In quantum dynamics the evolution of an observable takes the form
\begin{equation}
A(t) = U(t)^{-1} A  \, U(t)
,
\end{equation}
where $U(t)$ denotes the evolution operator from time zero to time $t$. Since
this is a similarity transformation, the spectrum of the observable is
unchanged during the evolution, so the range (in the sense of spectrum) is
preserved. 

In the classical case, the range of an observable $A$ is the range of the
function $A(x,k)$, i.e., the set of values that this quantity can take when
$(x,k)$ is varied.  This range is preserved by the classical
evolution. Indeed, the evolved observable at time $t$, $A(x,k;t)$, is nothing
else than $A(x(t),k(t))$, where $x(t)$ and $k(t)$ are the evolutions of $x$
and $k$ (i.e., the solution of the Hamilton equations). Since that evolution
is a bijection in the phase space the range of $A$ is unchanged.  For the
discussion of the hybrid case below, it is interesting to see the same thing
from another point of view.  Let us assume for simplicity that the range at
time $t$ is $[a,\infty[$ and $A(t)$ attains the minimum value $a$ at
    $(x_0,k_0)$.  In order for the range to change at $a$ one must have
    $\partial_t A(x_0,k_0;t)$ different from zero, however the classical
    equation is
\begin{equation}
\partial_t A(x,k;t) = \vnabla_x A(x,k;t) \cdot \vnabla_k H(x,k;t) -
\vnabla_k A(x,k;t) \cdot \vnabla_x H(x,k;t) 
.
\end{equation}
So $\vnabla_x A$ and $\vnabla_k A$ vanish at the minimum $(x_0,k_0)$ and so
does $\partial_t A(x,k;t)$.

The reason for the preservation of the range, either in the classical or the
quantum cases, is clear. The range of the observable is the set of all
possible values it can take in any of the states of the system. The evolution
corresponds to a change of the state of the system, but a change of state
cannot modify the range since it already accounts for all possible values.
In this view it seems natural to demand the preservation of the range of
observables during evolution also in the hybrid case.

For a hybrid observable we define its range as the union of all quantum
spectra from all points of the classical phase space. Another equivalent
definition of the range of $A$ is the set of values $\esp{A}_\rho$ can take as
$\rho$ varies (restricted by the conditions of positivity and
normalization). This range must not change during the evolution. In
particular, if $A$ is positive (as an operator at all points $(x,k)$) this
property ought to be preserved. However, this is not true for our dynamical
hybrid bracket.

A simple counterexample (of preservation of positivity) follows from the
particle with classical position-momentum and quantum spin-$\frac{1}{2}$
considered in Sec. \ref{sec:3}. Let us consider a generic observable
$A=a+\vb\cdot\vsigma$ [\Eq{3.13}]. Such observable will be positive at $t=0$
provided its spectrum is positive at all $(\vx,\vk)$. Since the local spectrum
is $a\pm \|\vb\|$, positivity requires
\begin{equation}
a(\vx,\vk;0) \ge \|\vb(\vx,\vk;0)\|
\quad \forall \vx,\vk
.
\label{eq:6.3}
\end{equation}
For the dynamics $H= g\vL\cdot\vS$, the evolution is indicated in
\Eq{3.17}. There we can see in the equation for $\vb(t)$, that the first and
third terms are bounded components as they depend on $R_t $, however, the
second term depends linearly with $t$. This term, which needs not vanish, will
eventually dominate for large (positive or negative) $t$ and will spoil the
positivity condition $a(t) \ge \|\vb(t)\|$. This shows that positivity is
not preserved in general by the hybrid dynamics.

The previous argument would not directly apply to a dynamics like that in
\Eq{3.17b} (with phase space based on $\vL$ only), nevertheless, also in this
case one can find instances of observables which are positive at $t=0$ but not
at all later times.

\subsection{Schrodinger picture}

In the Schrodinger picture $\rho$ must be a positive operator at all points.
This guarantees that the expectation value of arbitrary positive operators are
positive. For instance, $\esp{(A-\esp{A}_\rho)^2}_\rho \ge 0$ ~(with $A$
Hermitian). And the positivity of $\rho$ must hold at all times, so it ought
to be preserved by the dynamical evolution. This is true in the classical and
quantum cases, but it is not guaranteed in the hybrid dynamics.

Since we have already seen that the positivity of an observable is not
preserved in the hybrid case it follows that the positivity of $\rho$ will
also be spoiled. Indeed,
\begin{equation}
\esp{A(t)}_\rho = \esp{A}_{\rho(t)}
\end{equation}
in the Heisenberg and in Schrodinger picture, respectively. If $A$ is positive
but $\esp{A(t)}_\rho$ evolves to negative values (with positive $\rho$), in the
Schrodinger picture this must correspond to a $\rho(t)$ that cannot be
positive definite.

We can consider again the example of the particle with classical
position-momentum and quantum spin-$\frac{1}{2}$. In that setting
a Hamiltonian takes the form
\begin{equation}
H = h_c + \vh_q \cdot \vS
,
\end{equation}
where $h_c$ and $\vh_q$ are  $c$-number functions of $(\vx,\vk)$. The density
matrix is
\begin{equation}
\rho = \alpha + \vbeta \cdot \vS
,
\end{equation}
where $\alpha$ and $\vbeta$ are $c$-number functions of $(\vx,\vk)$ and $t$,
and the positivity condition on $\rho$ is $\alpha \ge \hbar \|\vbeta\|/2$.

The dynamical evolution, with the bracket of \Eq{5.24} for $n=2$, takes the
form
\begin{equation}
\frac{d\rho}{dt} = \php{H}{\rho}
=
\pc{h_c}{\alpha} + \pc{h_c}{\vbeta}\cdot\vS + 
\vh_q\times\vbeta\cdot\vS
+\frac{\hbar^2}{4} \pc{\vh_q\cdot}{\vbeta}
.
\end{equation}
That is
\begin{equation}\begin{split}
\partial_t \alpha &= 
\pc{h_c}{\alpha} + \frac{\hbar^2}{4} \pc{\vh_q\cdot}{\vbeta}
,\\
\partial_t \vbeta &= 
 \pc{h_c}{\vbeta} + \vh_q\times\vbeta
.
\end{split}\end{equation}

It is instructive to see how positivity of $\rho$ is preserved in the
classical and quantum cases. In the purely classical case $\vh_q=\vbeta=0$,
and the only remaining relation is $\partial_t \alpha = \pc{h_c}{\alpha}$.  If
$\alpha \ge 0$ ever reaches the value $0$ at some point, this must be a
minimum, and so $\pc{~}{\alpha}$ vanishes there; this prevents $\alpha$ from
passing from positive to negative values. On the other hand in the purely
quantal case, all functions are $(\vx,\vk)$-independent, and the non trivial
relations are $\alpha= \text{constant}$ and $\partial_t \vbeta =
\vh_q\times\vbeta$. This generates a continuous rotation of the vector
$\vbeta$ around $\vh_q$, so if $\hbar \|\vbeta\|/2 \le \alpha$ at $t=0$ it
will remain so at all times.

In the hybrid case positivity of $\rho$ is not so protected. Let us consider
for instance the simple case of 
\begin{equation}
\vh_q = h_q\hat{e}_z
,\qquad
 \vbeta = \beta\hat{e}_z,
\quad (t=0)
.
\end{equation}
With this dynamics the form of $\vbeta$ parallel to the $z$ axis is preserved
by the evolution. The equations become
\begin{equation}\begin{split}
\partial_t \alpha &= 
\pc{h_c}{\alpha} + \frac{\hbar^2}{4} \pc{h_q}{\beta}
,\\
\partial_t \beta &= 
 \pc{h_c}{\beta} 
.
\end{split}\end{equation}
Here we see that the positivity condition $\alpha \ge \hbar|\beta|/2$ at
$t=0$ needs not be preserved at later times: any of the functions
$\alpha^\prime \equiv \alpha \pm \hbar\beta/2$ fulfills the equation
\begin{equation}
\partial_t \alpha^\prime = 
\pc{h_c}{\alpha^\prime} + \frac{\hbar^2}{4} \pc{h_q}{\beta}
.
\end{equation}
If $\alpha^\prime$ becomes zero at some point $(\vx,\vk)$, this will be a
minimum and $\pc{h_c}{\alpha^\prime}$ will vanish there, however, the other
contribution $ \frac{\hbar^2}{4} \pc{h_q}{\beta}$ can be negative, since
$\beta$ follows its own autonomous dynamics. So nothing prevents from having
$\partial_t \alpha^\prime < 0 $ at a point where $\alpha^\prime=0$. In that
case $\alpha^\prime$ would become negative there, implying a violation of
positivity of $\rho(t)$. Of course, an analysis similar to this one can
carried out in the Heisenberg picture to show that positivity of an observable
is not preserved.

The explicit solution of $\rho(t)$ in \Eq{5.30} also illustrates that
positivity is not preserved in general: The term in $\alpha(t)$ depending
linearly with $t$ (while the other terms are bounded) implies that $\alpha(t)$
can become negative.

\subsection{Discussion}

Clearly, preservation of the positivity of $\rho$ is a condition that must be
imposed to any hybrid dynamics. Since our solution is unique using the strong
postulates, we have to consider weaker axioms. The condition of preservation
of positivity is somewhat more difficult to impose because it is does not have
an algebraic form. We do not attempt to work it out here.

We have seen that there is no problem of positivity in the classical or quantum
cases and this is also true in the Ehrenfest hybrid formalism discussed at the
end of Sec. \ref{sec:2c}. There, the density $\sigma(\vX)$ on $\cM$ evolves as
$\sigma(\vX;t) = \sigma(\vX(-t))$ and obviously positivity is preserved (the
points $\vX$ move but preserve their weight).  The $\rho$ corresponding to a
given $\sigma$ is easily obtained from
\begin{equation}
\esp{A}_\sigma  = \int dx \, dk \, d\psi \, d\psi^* 
\, \sigma(\vX) \esp{A(\vx,\vk)}_\psi 
= \Esp{A \rho}
,
\end{equation}
hence
\begin{equation}
\rho = \int  d\psi \, d\psi^* 
\, \sigma(\vX) | \psi\rangle \langle \psi |
\,.
\end{equation}
Note, incidentally, that there is an enormously larger number of different
$\sigma$ than $\rho$. This is a manifestation of the explosion of observables
in the Ehrenfest approach. Also, different $\sigma$ producing the same $\rho$
at $t=0$ will evolve producing different $\rho(t)$ at later times. This 
breaks statistical consistency.

Clearly $\rho$ is positive since $\sigma$ is positive, and this property is
preserved by the evolution.  We noted at the end of Sec. \ref{sec:2c} that our
formulation can be cast similarly on $\cM$ with the observables being
functions of bilinear type, but it cannot be based on evolution from orbits
$\vX(t)$ on $\cM$ as this breaks bilinearity. Nevertheless, one observes that
the positivity preserving flow of $\sigma(\vX)$ follows a continuity equation
$\partial_t \sigma = \vnabla_X\cdot (\vv_X \sigma)$ while the observables obey
the adjoint relation $\partial_t F = \vv_X \cdot \vnabla_X F$.\footnote{In the
  Hamiltonian case both operators $\vnabla_X \cdot \vv_X~$ and $\vv_X \cdot
  \vnabla_X ~$ coincide due to $(\vnabla_X\cdot \vv_X) = 0$.}  The
conservation properties of the continuity equation depend on the fact that
$\vv_X \cdot \vnabla_X$ is a derivation. If this would translate {\em as is}
to our formulation, we should require that, in order to preserve $\rho$, the
adjoint of $\php{~}{~}$ should be a derivation, that is $\ph{~}{~}$ should be
a derivation. If this is the case, we can certainly conclude that no solution
exists, as we showed above (Sec. \ref{sec:2b}) that $\ph{~}{~}$ can never be a
derivation in $\cA$ (even with the weak postulates). However the translation
is not completely obvious so we cannot give a final answer at present.

\section{Summary and Conclusions}
\label{sec:8}

In this work we have followed the seemingly most natural path to formulating
exact quantum-classical hybrid systems.  As discussed in Sec. \ref{sec:2},
this entails to have a set of hybrid observables for which we take the tensor
product of the classical and the quantum spaces of observables [\Eq{2.5}], in
full analogy with the classical-classical or quantum-quantum cases. Next, in
the Heisenberg picture we have to provide a dynamical bracket between
observables, similar to the Poisson bracket and the commutator,
[\Eq{2.7a}]. To do so we have introduced conditions to be fulfilled by the
bracket. The minimal postulates are those in Eqs. (\ref{eq:1.8}),
(\ref{eq:1.8a}) and (\ref{eq:2.12}), together with the Lie bracket property,
and in particular the Jacobi identity [\Eq{4.11}]. Another essential condition
is the preservation of the range of the observable (Sec. \ref{sec:6}). We have
not attempted to obtain the most general Lie bracket fulfilling the minimal
postulates as that problem looks hard. Instead, we have considered stronger
postulates [\Eq{1.19}] for which we have been able to obtain a complete
solution in the case of finite-dimensional Hilbert spaces, [\Eq{4.6}]. This
complements a previous result in \cite{Caro:1998us} showing that such
dynamical bracket does not exists for $\cH=L^2(\R^n)$.  The proof of the Lie
bracket property is given in Sec. \ref{sec:4b}.  The bracket is obtained
constructively in Sec. \ref{sec:7} where it is shown that the solution is
unique (with some additional assumptions of universality).  Generalizations
are discussed in Appendix \ref{app:A}.  The adjoint bracket, which describes
the evolution of the density matrix in the Schrodinger picture is explicitly
obtained in Sec. \ref{sec:5}. The ideas have been illustrated in
Secs. \ref{sec:3} and \ref{sec:5} with the case of a particle with classical
position-momentum and quantum spin-$\frac{1}{2}$.  The failure to comply with
positivity (Sec. \ref{sec:6}) is a serious problem which invalidates the
solution found.  One possible way out would be to weaken the postulates. This
would allow new solutions perhaps one of them preserving the positivity of the
matrix density.  In the same section we have speculated that the preservation
of the positivity of the density matrix in its evolution could require
$\ph{~}{H}$ to be a derivation (as in the classical or quantum cases). Since
this requisite can never be satisfied in the hybrid case, this would be a
no-go situation. Apart from this speculation, there is the real possibility
that no acceptable solution exists even for the minimal postulates. In this
case, either one has to admit more exotic formulations somehow sorting the
many problems noted in Sec. \ref{sec:2c}, or else admit that no formulation of
the exact type exists for quantum-classical systems, which would remain as
useful approximations of quantum-quantum systems.

\begin{acknowledgments}
I thank M. Hall and M. Reginatto for bringing Ref.~\cite{Schiller:1962} to my
attention.  This work was supported by Spanish Ministerio de Econom\'{\i}a y
Competitividad and European FEDER funds (grant FIS2014-59386-P), and by the
Agencia de Innovaci\'on y Desarrollo de Andaluc\'{\i}a (grant FQM225).
\end{acknowledgments}

\appendix

\section{Mathematical structure of the hybrid bracket}
\label{app:A}

Here we want to analyze the mathematical structure of our hybrid bracket and
its possible generalizations.

In the theory of Lie algebras, the semidirect sum of two Lie algebras is
defined as follows \cite{Barut:1986dd}. Let $L_1$ and $L_2$ be two Lie
algebras with Lie products denoted by $[~,~]_1$ and $[~,~]_2$, respectively.
In addition, for any $X \in L_2$, let $D_X$ be a derivation on $L_1$, that is,
\begin{equation}
\forall X \in L_2 \quad
\forall Y,Z \in L_1 \quad
D_X(Y) \in L_1,\qquad
D_X([Y,Z]_1) = [D_X(Y),Z]_1+[Y,D_X(Z)]_1
.
\label{eq:B1}
\end{equation}
Then, if $X \to D_X$ is a homomorphism, that is,
\begin{equation}
\forall X,Y\in L_2, \quad
D_{[X,Y]_2} = [D_X,D_Y]
,
\label{eq:B2}
\end{equation}
one can define a new Lie algebra 
\begin{equation}
L = L_1 \oplus_s L_2
\end{equation}
called the semidirect sum of $L_1$ and $L_2$. The vector space of $L$ is the
{\em external} direct sum of $L_1$ and $L_2$. External means that the elements
of $L$ are ordered pairs of elements  $X_1\in L_1$ and $X_2 \in L_2$ denoted
by $X_1+X_2$.  Within $L$, $L_1$ and $L_2$ are linearly independent by
definition and in the finite dimensional case $\dim L=\dim L_1+\dim L_2$.
The Lie product of two elements of $L$ is defined as
\begin{equation}
[X_1+X_2,Y_1+Y_2] = [X_1,Y_1]_1 + [X_2,Y_2]_2 + D_{X_2}(Y_1) - D_{Y_2}(X_1) ,
\end{equation}
where $X_1, Y_1\in L_1$ and $X_2, Y_2\in L_2$.  It is readily verified that
this is a Lie bracket using Eqs. (\ref{eq:B1}) and (\ref{eq:B2}).

In the hybrid quantum-classical case, $L_2$ is $\cA_c$, understood as a set of
classical observables endowed with the Poisson bracket as Lie product. This
set needs not be all classical observables, but it must be a Lie algebra,
i.e., a linear space closed under the Poisson bracket
\begin{equation}
\pc{\cA_c}{\cA_c} \subseteq \cA_c .
\label{eq:B5}
\end{equation}

Further, $L_1$ is $\hcA$ understood as a set
of hybrid observables (functions on the phase space taking values in
operators) with the requirement of being a Lie algebra with the
commutator, that is, a linear space such that
\begin{equation}
\pq{\hcA}{\hcA} \subseteq \hcA
.
\label{eq:B6}
\end{equation}
So here we no longer insist on the Hilbert space being finite dimensional and
the observables of $\hcA$ being traceless.  Also, $\hcA$ needs not be closed
under multiplication. For the derivation $D_X$ we take the Poisson bracket,
$D_X = \pc{X}{~}$, since it fulfills Eqs. (\ref{eq:B1}) and
(\ref{eq:B2}). Because $D_X(Y)\in L_2$, we further require
\begin{equation}
\pc{\cA_c}{\hcA} \subseteq \hcA
.
\label{eq:B7}
\end{equation} 
Under these conditions we can construct the
semidirect sum
\begin{equation}
\cA = \hcA \oplus_s \cA_c
,
\end{equation}
with
\begin{equation}
\ph{\hA + C }{ \hA^\prime + C^\prime } = \pq{\hA}{\hA^\prime} +
\pc{C}{C^\prime} + \pc{C}{\hA^\prime} - \pc{C^\prime}{\hA} 
\end{equation}
which will be a Lie bracket in $\cA$.

One way to guarantee Eqs. (\ref{eq:B5}), (\ref{eq:B6}) and (\ref{eq:B7}), is
to include in $\cA_c$ all classical observables and take $\hcA = \cA_c \otimes
\hcA_q$, where $ \hcA_q$ is a Lie algebra of operators with the commutator
(i.e., $\pq{\hcA_q}{\hcA_q} \subseteq \hcA_q$).  In this case, $\cA = \hcA
\oplus \cA_c = (\cA_c\otimes\hcA_q) \oplus \cA_c$ as vector space, or
equivalently $\cA = \cA_c \otimes \cA_q$ with $\cA_q = \hcA_q \oplus U$, where
$U$ is a one dimensional vector space.

In the particular case in which the identity of the Hilbert space {\em is not}
contained in $\hcA_q$, one can simply identify $U$ with $\{\lambda \unit,
\lambda \in \R \}$, and $\hA + C$ with $\hA + C \unit$. This is what has been
done in Sec. \ref{sec:4} for a finite dimensional Hilbert, along with choice
of $\hcA_q$ as the set of traceless operators.

For $\cH= L^2(\R^n)$, the condition $\pq{\hcA_q}{\hcA_q} \subseteq \hcA_q$
implies that $\unit \in \hcA_q$ (at least for $\hcA_q$ sufficiently large to
be useful). So, if one insists in carrying out the construction above, one
would have $\unit$ in $\hcA_q$ and an additional, linearly independent
$\unit^\prime$ in $U$. The resulting $\cA$, or more specifically $\cA_q =
\hcA_q \oplus U$, does not conform to the structure assumed in
Sec. \ref{sec:2a}, since $\hcA_q$ would be all quantum operators and $\cA_q$
would be something else, with a new dimension along $\unit^\prime$ outside the
original Hilbert space and with no obvious physical meaning.

Another observation is that in the finite dimensional case, one can choose
some Lie algebra of operators $\hcA_q$ not including $\unit$ and make the
construction with $\cA_q = \hcA_q \oplus U$ (with $U=\{\lambda \unit, \lambda
\in \R \}$) and $\cA = \cA_c\otimes\cA_q$, as already noted. Often, for a
given space $\cA_q$, the requirement of $\hcA_q$ being a Lie algebra will
completely fix this space in the decomposition $\cA_q = \hcA_q \oplus U$
[always with $U=\{\lambda \unit, \lambda \in \R \}$], but not in all cases. In
those cases, each choice of $\hcA_q$ will define a different hybrid bracket.
So the proof of uniqueness in Sec. \ref{sec:7} relies on the assumption that
all operators are included in $\cA_q$. This assumption is also used explicitly
in the construction of the hybrid bracket in the Schrodinger picture,
$\php{~}{~}$.

\section{Proof of \Eq{7.19}}
\label{app:B}

In this Appendix we proof that Eqs (\ref{eq:7.14}) and (\ref{eq:7.18}) imply
\Eq{7.19}. Since, \Eq{7.18} is identical for the three brackets
$\bph{~}{~}_\delta$, $\bph{~}{~}_f$ and $\bph{~}{~}_d$, we just consider one of
them explicitly.

We explicitly assume that our classical sector is of position-momentum type,
with $\cA_c$ containing all functions of $(\vx,\vk)$. It is not clear whether
the proof covers the case of restricted subsets of classical observables (but
still forming a Lie algebra, $\pc{\cA_c}{\cA_c} \subseteq \cA_c$) as in
\Eq{2.3}.

Following \cite{Caro:1998us}, let us introduce a plane-wave basis in the set
of functions defined on the phase space
\begin{equation}
e_r = \exp(i \vk_r \cdot \vx - i \vx_r \cdot \vk )
,\qquad
\vx_r, \vk_r \in \R^n
.
\end{equation}
$(\vx,\vk)$ are the classical dynamical variables and $\vx_r, \vk_r $
parameters. These basis functions fulfill the relations
\begin{equation}
\pc{e_r}{\vx} = i \vx_r \, e_r
,\qquad
\pc{e_r}{\vk} = i \vk_r \, e_r
,
\label{eq:A2}
\end{equation}
and also
\begin{equation}
\pc{e_r}{e_s} = v_{rs} \, e_{r+s}
\end{equation}
with
\begin{equation}
v_{rs} \equiv \vk_r\cdot \vx_s -  \vx_r\cdot \vk_s
,
\qquad
(\vx_{r+s},\vk_{r+s}) \equiv (\vx_r+\vx_s,\vk_r+\vk_s)
.
\end{equation}

Using this basis
\begin{equation}
\bph{e_r}{e_s}_\delta = \sum_t F_{rst} e_t
,
\label{eq:A5}
\end{equation}
for some function $F_{rst}$. However, using Eqs. (\ref{eq:7.18}) and
(\ref{eq:A2}),
\begin{equation}
\pc{ \bph{e_r}{e_s}_\delta }{ \vx } =  i\vx_{r+s} \bph{e_r}{e_s}_\delta
,\quad
\pc{ \bph{e_r}{e_s}_\delta }{ \vk } =  i\vk_{r+s} \bph{e_r}{e_s}_\delta
.
\end{equation}
This implies [comparing with \Eq{A2} ] that in the sum over $t$ in 
\Eq{A5}, only the term $t=r+s$ can have a contribution. That is,
\begin{equation}
\bph{e_r}{e_s}_\delta = F_{rs} e_{r+s}
,
\label{eq:A5a}
\end{equation}
where $F_{rs}$ is some function of $(\vx_r,\vk_r)$ and
$(\vx_s,\vk_s)$. 

In order to pinpoint this function, let us impose \Eq{7.18} with
$C=e_r$, $C^\prime= e_s$ and $C^{\prime\prime} = e_t$. This gives
\begin{equation}
F_{rs} v_{r+s,t} = 
 v_{rt} F_{r+t,s} + v_{st} F_{r,s+t}
\quad \forall r,s,t
.
\label{eq:A8}
\end{equation}
Let us show that this relation implies that $F_{rs}$ must be a function of
$v_{rs}$ only.\footnote{The idea is that if $t$ is such that $v_{st}=0$ then
  $v_{r+s,t} = v_{rt}$ and hence $F_{rs}=F_{r+t,s}$. That is, $r$, which is a
  $2n$-dimensional variable, can be changed in $2n-1$ directions (those of $t$
  restricted by $v_{st}=0$) without modifying $F_{rs}$, so the dependence on
  $r$ is really through $v_{rs}$.} To do this let us take
\begin{equation}
\vk_t = -\vk_r ,\qquad
\vx_t = -\vx_r - v_{rs} \frac{\vk_s}{\vk_s^2}
.
\end{equation}
With this choice
\begin{equation}
v_{st}=0, \qquad
v_{r+s,t} = v_{rt}
.
\end{equation}
Hence $F_{rs}=F_{r+t,s}$, or more explicitly
\begin{equation}
F(\vx_r,\vk_r;\vx_s,\vk_s) = 
F(-v_{rs}\frac{\vk_s}{\vk_s^2},\bm{0};\vx_s,\vk_s) 
.
\end{equation}
That is, for given $(\vx_s,\vk_s)$, all the dependence on $(\vx_r,\vk_r)$ is
through $v_{rs}$. Taking an analogous choice of $t$ this time with $v_{rt}=0$,
one obtains that all dependence on $(\vx_s,\vk_s)$ is also through
$v_{rs}$. So
\begin{equation}
F_{rs} = f(v_{rs})
,
\end{equation}
for some function $f(v)$ to be specified, and \Eq{A8} becomes
\begin{equation}
f(v_{rs}) v_{r+s,t} = 
 v_{rt} f(v_{r+t,s}) + v_{st} f(v_{r,s+t})
.
\label{eq:A13}
\end{equation}
Taking
\begin{equation}
v_1 \equiv v_{rs}
,\qquad
v_2 \equiv v_{r+t,s}
,\qquad
v_3 \equiv v_{r,s+t}
,
\end{equation}
as independent variables, so that
\begin{equation}
v_{r+s,t} = v_3-v_2
,\qquad
v_{rt}= v_3-v_1
,\qquad
v_{st} =v_1 - v_2
,
\end{equation}
the equation on $f(v)$ becomes
\begin{equation}
f(v_1) (v_3-v_2) = (v_3-v_1) f(v_2) + (v_1-v_2) f(v_3)
,
\label{eq:A16}
\end{equation}
where $v_1,v_2,v_3$ are arbitrary.  Paying attention to any of the three
variables with the other two fixed, it follows that
\begin{equation}
f(v) = \alpha v + \beta
,
\end{equation}
for some constants $\alpha$ and $\beta$. In turn this forms fulfills \Eq{A16}
identically. Hence the most general solution is
\begin{equation}
\bph{e_r}{e_s}_\delta = (\alpha v_{rs} + \beta )e_{r+s}
,
\end{equation}
which corresponds to the bracket
\begin{equation}
\bph{C}{C^\prime}_\delta = \alpha \pc{C}{C^\prime} + \beta C C^\prime
.
\end{equation}
Since $\bph{~}{~}_\delta$ and $\bph{~}{~}_d$ are antisymmetric, $\beta=0$ in
those cases. On the other hand $\bph{~}{~}_f$ is symmetric, so $\alpha=0$ for
that bracket. This proves \Eq{7.19}.


\begin{thebibliography}{99}

\bibitem{Bohr:1958}
  N.~Bohr, {\it Atomic Physics and Human Knowledge} (Wiley, New York, 1958).

\bibitem{Heisenberg:1958}
  W.~Heisenberg, {\it Physics and Philosophy: The Revolution in
  Modern Science} (Harper Perennial Modern Classics, London, 2007).

\bibitem{dEspagnat:1976}
  B.~ d’Espagnat, {\it Conceptual Foundations of Quantum Mechanics}
  (Addison Wesley, Reading, 1976).

\bibitem{Allahverdyan:2011cx} 
  A.~E.~Allahverdyan, R.~Balian and T.~M.~Nieuwenhuizen,
  ``Understanding quantum measurement from the solution of dynamical models,''
  Phys.\ Rept.\  {\bf 525}, 1 (2013)
  doi:10.1016/j.physrep.2012.11.001
  [arXiv:1107.2138 [quant-ph]].


\bibitem{Rosenfeld:1963}
  L.~Rosenfeld,
  ``On quantization of fields,''
  Nucl. Phys. {\bf 40} (1963) 353.
  doi:10.1016/0029-5582(63)90279-7

\bibitem{Kibble:1979jn} 
  T.~W.~B.~Kibble and S.~Randjbar-Daemi,
  ``Nonlinear Coupling of Quantum Theory and Classical Gravity,''
  J.\ Phys.\ A {\bf 13}, 141 (1980).
  doi:10.1088/0305-4470/13/1/015

\bibitem{Isham:1980sb}
 C. J. Isham, in {\it Quantum Gravity: An Overview, Quantum
Gravity 2: a Second Oxford Symposium}, edited by C. J. Isham,
R. Penrose, and D. W. Sciama (Oxford University Press, Oxford,
1981), pp. 1–62.

\bibitem{Page:1981aj} 
  D.~N.~Page and C.~D.~Geilker,
  ``Indirect Evidence for Quantum Gravity,''
  Phys.\ Rev.\ Lett.\  {\bf 47}, 979 (1981).
  doi:10.1103/PhysRevLett.47.979

\bibitem{Alvarez:1988tb} 
  E.~Alvarez,
  ``Quantum Gravity: A Pedagogical Introduction To Some Recent Results,''
  Rev.\ Mod.\ Phys.\  {\bf 61}, 561 (1989).
  doi:10.1103/RevModPhys.61.561

\bibitem{Boughn:2008jx} 
  S.~Boughn,
  ``Nonquantum Gravity,''
  Found.\ Phys.\  {\bf 39}, 331 (2009).
  doi:10.1007/s10701-009-9282-0
  [arXiv:0809.4218 [gr-qc]].

\bibitem{Carlip:2008zf} 
  S.~Carlip,
  ``Is Quantum Gravity Necessary?,''
  Class.\ Quant.\ Grav.\  {\bf 25}, 154010 (2008).
  doi:10.1088/0264-9381/25/15/154010
  [arXiv:0803.3456 [gr-qc]].

\bibitem{Derakhshani:2013qya} 
  M.~Derakhshani,
  ``Newtonian Semiclassical Gravity in the Ghirardi-Rimini-Weber Theory with Matter Density Ontology,''
  Phys.\ Lett.\ A {\bf 378}, 990 (2014)
  doi:10.1016/j.physleta.2014.02.005
  [arXiv:1304.0471 [gr-qc]].

\bibitem{Aqvist:1993}
  J.~\r{A}qvist and A.~ Warshel,
  ``Simulation of enzyme reactions using valence bond force fields and other
  hybrid quantum/classical approaches,''
  Chem. Rev. {\bf 93}, 2523  (1993).
  doi:10.1021/cr00023a010

\bibitem{Monard:1996}
  G.~Monard, M.~Loos, V.~Th\'ery, K.~Baka, and J-L.~Rivail,
  ``Hybrid classical quantum force field for modeling very large molecules,''
  Int. J. Quantum Chem. {\bf 58}, 153 (1996).
  doi:10.1002/(SICI)1097-461X(1996)58:2$<$153::AID-QUA4$>$3.0.CO;2-X

\bibitem{Bornemann:1996}
  F.~A.~Bornemann, P.~Nettesheim, and C.~ Sch\"utte
  ``Quantum‐classical molecular dynamics as an approximation to full quantum dynamics''
  J. Chem. Phys. {\bf 105}, 1074 (1996).
  doi:10.1063/1.471952

\bibitem{Prezhdo:1997}
  O.~V.~Prezhdo and P.~J.~Rossky,
  ``Mean-field molecular dynamics with surface hopping,''
  J. Chem. Phys. {\bf 107}, 825 (1997).
  doi:10.1063/1.474382

\bibitem{Kapral:1999}
  R.~Kapral and G.~Ciccotti,
  ``Mixed quantum-classical dynamics,''
  J. Chem. Phys. {\bf 110}, 8919 (1999).
  doi:10.1063/1.478811

\bibitem{Csanyi:2004}
  G.~Cs\'anyi, T.~Albaret, M.~C.~Payne and A.~De Vita,
  ``Learn on the Fly”: A Hybrid Classical and Quantum-Mechanical Molecular Dynamics Simulation,''
  Phys. Rev. Lett. {\bf 93}, 175503 (2004).
  doi:10.1103/PhysRevLett.93.175503

\bibitem{Magnus:1954zz}
  W.~Magnus,
  ``On the exponential solution of differential equations for a linear operator,''
  Commun.\ Pure Appl.\ Math.\ {\bf 7}, 649 (1954).
  doi:10.1002/cpa.3160070404

\bibitem{Blanes:2009}
  S.~Blanes, F.~Casas, J.~A.~Oteo, J.~Ros,
  ``The Magnus expansion and some of its applications,''
  Phys. Rep. {\bf 470}, 151 (2009).
  doi:10.1016/j.physrep.2008.11.001

\bibitem{Barcelo:2012ja} 
  C.~Barcelo, R.~Carballo-Rubio, L.~J.~Garay and R.~Gomez-Escalante,
  ``Hybrid classical-quantum formulations ask for hybrid notions,''
  Phys.\ Rev.\ A {\bf 86}, 042120 (2012)
  doi:10.1103/PhysRevA.86.042120
  [arXiv:1206.7036 [quant-ph]].

\bibitem{Aleksandrov:1981}
  I.~V. Aleksandrov,
  ``The statistical dynamics of a system consisting of a classical and a quantum  subsystem,''
  Z. Naturforsch. {\bf 36A}, 902 (1981).

\bibitem{Boucher:1988ua} 
  W.~Boucher and J.~H.~Traschen,
  ``Semiclassical Physics and Quantum Fluctuations,''
  Phys.\ Rev.\ D {\bf 37}, 3522 (1988).
  doi:10.1103/PhysRevD.37.3522

\bibitem{Anderson:1994si} 
  A.~Anderson,
  ``Quantum back reaction on 'classical' variables,''
  Phys.\ Rev.\ Lett.\  {\bf 74}, 621 (1995).
  doi:10.1103/PhysRevLett.74.621
  [hep-th/9406182].

\bibitem{Caro:1998us} 
  J.~Caro and L.~L.~Salcedo,
  ``Impediments to mixing classical and quantum dynamics,''
  Phys.\ Rev.\ A {\bf 60}, 842 (1999)
  doi:10.1103/PhysRevA.60.842
  [quant-ph/9812046].

\bibitem{Goldstein:2002}
  H.~Goldstein, C.~Poole, and J.~Safko, {\it Classical Mechanics, 3rd ed.}
  (Addison Wesley, New York, 2002).

\bibitem{Oliynyk:2016jwt} 
  T.~A.~Oliynyk,
  ``Classical-Quantum Limits,''
  Found.\ Phys.\  {\bf 46}, no. 12, 1551 (2016)
  doi:10.1007/s10701-016-0028-5
  [arXiv:1508.04477 [quant-ph]].

\bibitem{Prezhdo:2006}
O.~V. Prezhdo,
  ``A quantum-classical bracket that satisfies the Jacobi identity,''
  J. Chem. Phys. {\bf 124}, 201104 (2006).
  doi:10.1063/1.2200342

\bibitem{Salcedo:2007zk} 
  L.~L.~Salcedo,
  ``Comment on `A quantum-classical bracket that satisfies the Jacobi identity' [J.Chem.Phys.124,201104 (2006)],''
  J.\ Chem.\ Phys.\  {\bf 126}, 057101 (2007)
  doi:10.1063/1.2431650
  [quant-ph/0701054].

\bibitem{Prezhdo:2007}
O.~V. Prezhdo,
``Reply to “Comment on ‘A quantum-classical bracket that satisfies the Jacobi identity’ ” [J. Chem. Phys.124, 201104 (2006)],''
  J.\ Chem.\ Phys.\  {\bf 126}, 057102 (2007).
  doi:10.1063/1.2431651

\bibitem{Schwinger:1970xc} 
  J.~Schwinger,
  ``Particles, sources, and fields. Volume 1,''
  Reading, Mass., 1970, 425p.

\bibitem{Salcedo:1995jr} 
  L.~L.~Salcedo,
  ``Absence of classical and quantum mixing,''
  Phys.\ Rev.\ A {\bf 54}, 3657 (1996)
  doi:10.1103/PhysRevA.54.3657
  [hep-th/9509089].

\bibitem{Sahoo:2004}
  D.~Sahoo,
  ``Mixing quantum and classical mechanics and uniqueness of Planck{'}s constant,''
  J. Phys. A {\bf 37}, 997 (2004).
  doi:10.1088/0305-4470/37/3/031

\bibitem{Dass:2009qu} 
  T.~Dass,
  ``A Stepwise Planned Approach to the Solution of Hilbert's Sixth Problem. I : Noncommutative Symplectic Geometry and Hamiltonian Mechanics,''
  arXiv:0909.4606 [math-ph].

\bibitem{Negele:1988vy} 
  J.~W.~Negele and H.~Orland,
  ``Quantum Many Particle Systems,''
  Redwood City, USA: Addison-Wesley (1988) 459 P. (Frontiers in Physics, 68)

\bibitem{Diosi:1999py} 
  L.~Diosi, N.~Gisin and W.~T.~Strunz,
  ``Quantum approach to coupling classical and quantum dynamics,''
  Phys.\ Rev.\ A {\bf 61}, 22108 (2000)
  doi:10.1103/PhysRevA.61.22108, 10.1103/PhysRevA.61.022108
  [quant-ph/9902069].

\bibitem{Craig:2005}
  C.~F. Craig, W.~R. Duncan and O.~V. Prezhdo,
  ``Trajectory Surface Hopping in the Time-Dependent Kohn-Sham Approach for Electron-Nuclear Dynamics,''
  Phys.Rev.Lett. {\bf 95}, 163001 (2005).
  doi:10.1103/PhysRevLett.95.163001

\bibitem{Zhang:2006} 
Qi Zhang and Biao Wu,
  ``General Approach to Quantum-Classical Hybrid Systems and Geometric Forces,''
  Phys.\ Rev.\ Lett. {\bf 97}, 190401 (2006)
  doi:10.1103/PhysRevLett.97.190401

\bibitem{Alonso:2011}
  J.~L.~Alonso, A.~Castro, J.~Clemente-Gallardo, J.~C.~Cuchi, P.~Echenique and F.~Falceto,
  ``Statistics and Nos{\'e} formalism for Ehrenfest dynamics,''
  J. Math. Phys. {\bf 44}, 395004 (2011).
  doi:10.1088/1751-8113/44/39/395004

\bibitem{Elze:2011hi} 
  H.~T.~Elze,
  ``Linear dynamics of quantum-classical hybrids,''
  Phys.\ Rev.\ A {\bf 85}, 052109 (2012)
  doi:10.1103/PhysRevA.85.052109
  [arXiv:1111.2276 [quant-ph]].

\bibitem{Diosi:1997mt} 
  L.~Diosi and J.~J.~Halliwell,
  ``Coupling classical and quantum variables using continuous quantum measurement theory,''
  Phys.\ Rev.\ Lett.\  {\bf 81}, 2846 (1998)
  doi:10.1103/PhysRevLett.81.2846
  [quant-ph/9705008].

\bibitem{Salcedo:2012qx} 
  L.~L.~Salcedo,
  ``Statistical consistency of quantum-classical hybrids,''
  Phys.\ Rev.\ A {\bf 85}, 022127 (2012)
  doi:10.1103/PhysRevA.85.022127
  [arXiv:1201.4237 [quant-ph]].

\bibitem{Hall:2005ax} 
  M.~J.~W.~Hall and M.~Reginatto,
  ``Interacting classical and quantum ensembles,''
  Phys.\ Rev.\ A {\bf 72}, 062109 (2005)
  doi:10.1103/PhysRevA.72.062109
  [quant-ph/0509134].

\bibitem{Hall:2008}
  M.~J.~W.~Hall,
  ``Consistent classical and quantum mixed dynamics,''
  Phys.\ Rev.\ A {\bf 78}, 042104 (2008).
  doi:10.1103/PhysRevA.78.042104

\bibitem{Sudarshan:1976bt} 
  G.~Sudarshan,
  ``Interaction between classical and quantum systems and the measurement of quantum observables,''
  Pramana {\bf 6}, 117 (1976).
  doi:10.1007/BF02847120

\bibitem{Chruscinski:2011}
  D.~Chru\'sci\'nski, A.~Kossakowski, G.~Marmo and E. C. G. Sudarshan,
  ``Dynamics of Interacting Classical and Quantum Systems,''
  Open Syst. Inf. Dyn. {\bf 18}, 339 (2011).
  doi:10.1142/S1230161211000236


\bibitem{Prezhdo:2001}
  O.~V. Prezhdo and C.~Brooksby,
  ``Quantum Backreaction through the Bohmian Particle,''
  Phys. Rev. Lett. {\bf 86}, 3215 (2001).
  doi:10.1103/PhysRevLett.86.3215

\bibitem{Salcedo:2003mp} 
  L.~L.~Salcedo,
  ``Comment on `Quantum back reaction through the Bohmian particle',''
  Phys.\ Rev.\ Lett.\  {\bf 90}, 118901 (2003).
  doi:10.1103/PhysRevLett.90.118901

\bibitem{Prezhdo:2003}
  O.~V.~Prezhdo and C.~Brooksby,
  ``Prezhdo and Brooksby Reply:,''
  Phys.\ Rev.\ Lett.\  {\bf 90}, 118902 (2003).
  doi:10.1103/PhysRevLett.90.118902

\bibitem{Radonjic:2012}
  M.~Radonji\'c, S.~Prvanovi\'c, and N.~Buri\'c,
  ``Hybrid quantum-classical models as constrained quantum systems''
  Phys.\ Rev.\ A {\bf 85}, 064101 (2012).
  doi:10.1103/PhysRevA.85.064101

\bibitem{Peres:2001} 
  A.~Peres and D.~R.~Terno,
  ``Hybrid classical-quantum dynamics,''
  Phys.\ Rev.\ A {\bf 63}, 022101 (2001).
  doi:10.1103/PhysRevA.63.022101

\bibitem{Schiller:1962}
  R.~Schiller,
  ``Quasi-Classical Theory of the Spinning Electron,''
  Phys.\ Rev. {\bf 125}, 1116 (1962).
  doi:10.1103/PhysRev.125.1116

\bibitem{Pascual:1984zb} 
  P.~Pascual and R.~Tarrach,
  ``Qcd: Renormalization For The Practitioner,''
  Lect.\ Notes Phys.\  {\bf 194}, 1 (1984).

\bibitem{Barut:1986dd} 
  A.~O.~Barut and R.~Raczka,
  ``Theory Of Group Representations And Applications,''
  Singapore, Singapore: World Scientific ( 1986) 717p.

\bibitem{Ahmadzadega:2016}
A.~Ahmadzadegan, R.~B.~Mann, D.~R.~Terno,
  ``How classical is a quantum oscillator?,''
  Phys.\ Rev.\ A {\bf 93}, 032122 (2016).
  doi:10.1103/PhysRevA.93.032122




\end{thebibliography}
\end{document}